\documentclass[prd,twocolumn,showpacs,nofootinbib]{revtex4}
\usepackage{times}
\usepackage{natbib}
\usepackage{epsfig}
\usepackage{amsmath}
\usepackage{amssymb}
\usepackage{bm}

\newcommand{\bdv}[1]{{\bf #1}}

\renewcommand{\AA}{\mathcal{A}}
\newcommand{\UU}{u}           
\newcommand{\dUU}{\delta u^0}
\newcommand{\VV}{V}         
\newcommand{\SVV}{v}       
\newcommand{\VVV}{v}      
\newcommand{\cc}{\lambda}
\newcommand{\VBB}{B}    
\newcommand{\VCC}{C}   
\newcommand{\TCC}{C}  
\newcommand{\dtm}{\delta^{t_p}_m}
\newcommand{\LL}{\mathcal{L}}  
\newcommand{\TT}{\mathcal{T}} 
\newcommand{\pr}{h}          
\newcommand{\gI}{I}         
\newcommand{\gII}{II}      
\newcommand{\gIII}{III}   
\newcommand{\DD}{\bm{\Psi}}   
\newcommand{\tp}{\tau}       
\newcommand{\up}[1]{{\rm #1}}

\newcommand{\beeq}{\begin{equation}}
\newcommand{\eneq}{\end{equation}}
\newcommand{\bear}{\begin{eqnarray}}
\newcommand{\enar}{\end{eqnarray}}
\newcommand{\RA}{\rightarrow}
\newcommand{\OO}{\mathcal{O}}
\newcommand{\kvec}{\bdv{k}}
\newcommand{\qvec}{\bdv{q}}
\newcommand{\vvec}{\bdv{v}}
\newcommand{\xvec}{\bdv{x}}
\newcommand{\HH}{\mathcal{H}}  
\newcommand{\gbar}{\bar g}    
\newcommand{\BB}{\mathcal{B}}
\newcommand{\CC}{\mathcal{C}} 
\newcommand{\px}{\varphi_{\chi}}
\newcommand{\pv}{\Psi}         

\begin{document}

\title{Proper-time hypersurface of non-relativistic matter flows: 
Galaxy bias in general relativity}

\author{Jaiyul Yoo$^{1,2}$}
\altaffiliation{jyoo@physik.uzh.ch}
\affiliation{$^1$Center for Theoretical Astrophysics and Cosmology,
Institute for Computational Science, University of Z\"urich}
\affiliation{$^2$Physics Institute, University of Z\"urich, 
Winterthurerstrasse 190, CH-8057, Z\"urich, Switzerland}

\begin{abstract}
We compute the second-order density fluctuation in the proper-time 
hypersurface of non-relativistic matter flows and relate it to the galaxy 
number density fluctuation, providing physical grounds for galaxy bias
in the context of general relativity. At the linear order,
the density fluctuation in the proper-time hypersurface
is equivalent to the density fluctuation in the comoving synchronous gauge,
in which two separate gauge conditions coincide.
However, at the second order, the density fluctuations 
in these gauge conditions differ, while both gauge conditions represent the 
same proper-time hypersurface. Compared to the density fluctuation in the 
temporal comoving and the spatial C-gauge conditions,
the density fluctuation in the 
commonly used gauge condition ($N=1$ and $N^\alpha=0$) 
violates the mass conservation at the second order.
We provide their physical interpretations in each gauge condition
by solving the geodesic equation and the nonlinear evolution equations
of non-relativistic matter. 
We apply this finding to the second-order
galaxy biasing in general relativity, which complements the second-order
relativistic description of galaxy clustering in Yoo \& Zaldarriaga (2014).
\end{abstract}


\maketitle

\section{Introduction}
\label{sec:intro}

The discovery of the late-time cosmic acceleration has created the problem
of the century in physics, spurring numerous theoretical and observational
investigations in the last decades. In particular, enormous amount of efforts
have been devoted to large-scale galaxy surveys that can be used to map the 
three-dimensional matter distribution. 
Millions of galaxies at higher redshift with larger sky coverage will be
measured in the upcoming future surveys such as the Dark Energy Spectroscopic
Instrument, the Large Synoptic Survey Telescope, and two space-based missions
Euclid and the Wide-Field Infrared Survey Telescope.

In light of this recent development in large-scale galaxy surveys,
the general relativistic description of galaxy clustering has been developed 
\cite{YOFIZA09,YOO10,CHLE11,BODU11,JESCHI12,YOHAET12}.
In the standard Newtonian description, gravity is felt
instantaneously across the horizon, and a hypersurface of simultaneity is
well defined. However, none of these are valid in general relativity, and
the Newtonian description breaks down 
on cosmological scales, in which dark energy models manifest themselves
or modified gravity
theories deviate from general relativity. The relativistic description is,
therefore, an indispensable tool in the era of precision cosmology.

Cosmological observations are performed by measuring photons emitted from 
distant sources like galaxies, and they are affected by the matter fluctuation
and the gravitational perturbations along the path to reach us. 
Of significant interest is, therefore, to derive the relation of the 
observable quantities to the physical quantities of sources 
and to understand
how various relativistic effects such as gravitational potential and curvature
perturbation affect this relation along
the light propagation. In this way, the relativistic
description of galaxy clustering naturally resolves gauge issues 
\cite{YOFIZA09,YOO10} that
often plague theoretical predictions, providing a complete description
\cite{YOO09}
of all the effects in galaxy clustering such as the redshift-space distortion,
the gravitational lensing, the integrated Sachs-Wolfe effect, and so on
 (see \cite{YOO14a} for a review).

A significant portion of the relativistic effects in galaxy clustering arise
due to the mismatch between the physical quantities 
and the observable quantities, and this mismatch is tackled
by tracing the light propagation backward in time and by deriving the their
relations. However, what we measure is galaxies, not matter, 
and the relation between the galaxy and the underlying
matter distributions, known as {\it galaxy bias}, is another difficulty in 
formulating the relativistic description of galaxy clustering.
In general, the physical galaxy number density can be separated into 
the mean $\bar n_g(\tp)$ and the fluctuation $\delta_g^\up{int}$ around it:
\beeq
\label{eq:stg}
n_g=\bar n_g(\tp)(1+\delta_g^\up{int})~,
\eneq
and the linear bias model \cite{KAISE84} in Newtonian dynamics shows that
the intrinsic galaxy fluctuation should be proportional to the matter
density fluctuation~$\delta_m$ on sufficiently large scales:
\beeq
\label{eq:biasN}
\delta_g^\up{int}=b~\delta_m~,
\eneq
where the proportionality constant~$b$ is the bias factor.
Since the separation of the mean and the fluctuation
in Eq.~(\ref{eq:stg}) is arbitrary and relies on
unspecified coordinate time~$\tp$, the biasing relation in Eq.~(\ref{eq:biasN})
makes little sense in the context of general relativity and
is gauge-dependent.

From the relativistic perspective,
\citet*{YOFIZA09} assumed that the galaxy number density is a function
of the matter density $\rho_m$ (not the matter density fluctuation~$\delta_m$)
at the same spacetime point:
\beeq
\label{eq:bb}
n_g=F[\rho_m]~.
\eneq
The galaxy biasing in \citet{BODU11} and
\citet{BRCRET12} is neglected or assumed to follow the matter density,
respectively (hence it is essentially equivalent to Eq.~[\ref{eq:bb}]
with $F$ being the identity function).
While this biasing scheme is fully general and covariant, it is physically
restrictive as the time evolution of galaxy number density is strictly
driven by the matter density evolution $\bar n_g\propto(1+z)^3$.
To relax this restriction, while
keeping the locality, additional freedom was provided in \citet{YOHAET12} 
to allow galaxy number density to depend on its local history 
(or {\it proper-time}), describing different evolutionary tracks of
galaxy number densities at the same matter density.

By arguing that the galaxy number density is a Newtonian gauge quantity
and its Poisson equation is related to the matter density fluctuation
$\delta_m^\up{syn}$ in the synchronous gauge,
\citet{CHLE11} chose the synchronous gauge for galaxy bias in general 
relativity, and the biasing relation in Eq.~(\ref{eq:biasN}) becomes
$\delta_g^\up{int}=b~\delta_m^\up{syn}$. \citet{JESCHI12} advocated
the constant-age hypersurface (or the proper-time hypersurface) for the
biasing relation, as the proper-time is the {\it only} locally measurable
quantity that carries physical significance
on large scales. A proper generalization in the context
of general relativity is made in \citet{BASEET11} by constructing a local Fermi
coordinate, in which local observables can be explicitly written in terms
of the local curvature and the local expansion rate.
These biasing schemes based on the proper-time hypersurface 
\cite{YOHAET12,CHLE11,JESCHI12,BASEET11}
are all equivalent to each other at the linear order.

Given these theoretical developments, it is rather straightforward, albeit
lengthy, to extend the relativistic formalism to the second order in 
perturbation. 
The second-order perturbations are naturally smaller than the linear-order
perturbations. However, they do contain critical and invaluable 
information about the perturbation generation mechanism in the early Universe.
In the standard single-field inflationary model, the Universe is well described
by its nearly perfect Gaussianity on large scales, in which the power spectrum
contains the complete information. However, any models beyond the single-field
inflationary model have additional degrees-of-freedom, and these additional
fields couple to the curvature perturbations, leaving non-trivial
signatures manifest in higher-order statistics such as the bispectrum
(see, e.g., \cite{BAGR12,DIKAET09,BEMA09}).
Even in the standard single field model, gravity waves generate
non-trivial trispectrum in the curvature perturbations \cite{MALDA03}.
These unique signatures in the initial condition are generically
subtle and nonlinear relativistic effects, requiring proper relativistic
treatments beyond the linear-order in perturbations.
In this respect, the second-order relativistic 
description of galaxy clustering provides an essential tool to probe
the early Universe in large-scale galaxy surveys, 
and it was recently
formulated \cite{YOZA14,BEMACL14b,DIDUET14}.

\citet{BEMACL14b} advocate that the proper-time hypersurface
(or the rest-frame of baryons and dark matter) should be used for 
second-order galaxy bias, and they chose the matter density fluctuation 
$\delta_m^\up{\gII}$ in the comoving-time orthogonal gauge 
(see our gauge choice~\gII\ in Table~\ref{tab:gauge}) that becomes 
comoving-synchronous gauge for a presureless medium. 
In \citet{DIDUET14}, the galaxy number density is approximated as the matter
density, and the second-order galaxy biasing is left for future work.
In \citet{YOZA14}, the proper-time hypersurface is also advocated
for the second-order galaxy biasing scheme, but no specific choice of gauge 
condition is discussed for computing the matter density fluctuation $\dtm$
in the proper-time hypersurface at the second order. 

Here we 
provide the missing ingredient, completing the full second-order 
relativistic description in \cite{YOZA14}.
We compute the second-order matter density fluctuation $\dtm$
in the proper-time hypersurface of non-relativistic matter flows.
In particular, we focus on several
gauge choices summarized in Table~\ref{tab:gauge} in computing~$\dtm$.
Interestingly, these common gauge conditions provide different matter
density fluctuations at the second order, posing a critical question
in formulating galaxy bias in general relativity --- {\it which one and why?
any gauge issues?} We show that the matter density fluctuation in
 gauge choice~\gI\ in Table~\ref{tab:gauge} is the correct and physical
choice for the
matter density fluctuation in the proper-time hypersurface that can be
used for galaxy bias in general relativity at the second order.

Technical details of these gauge conditions at the second order
are extensively discussed in \citet{HWNO06b} (see also 
\cite{NOHW04,HWNO07a} and \cite{UGWA14} for different derivations).
Here we provide
physical interpretations of each gauge condition 
and discuss how they can be applied to 
second-order galaxy bias in general relativity. 
The organization of the paper is as follows. In Sec.~\ref{sec:flow}, we
present the basic formalism for computing the flow of non-relativistic matter
and derive the matter density fluctuation in the proper-time hypersurface
by solving the geodesic equation. Various observers are defined in
Sec.~\ref{ssec:obs}. Several gauge choices in Table~\ref{tab:gauge} 
and the gauge issues associated with them are discussed in 
Sec.~\ref{sec:gauge} in computing the matter density
fluctuation in the proper-time hypersurface. In Sec.~\ref{sec:non}, we present
the nonlinear evolution equations and derive their solutions for each
gauge choice. Gauge issues in the solutions and their physical interpretation
are discussed in Sec.~\ref{sec:issue} and Sec.~\ref{sec:phy}, respectively.
Finally, we summarize our finding and discuss the implications of our results
in Sec.~\ref{sec:discussion}. Two appendices summarize useful relations that 
are used in the paper. 

Throughout the paper spacetime indices are represented by Latin indices,
while spatial indices by Greek indices. Equations and variables in this paper
should be considered nonlinear, unless perturbation order
 is specifically mentioned.

\section{Flow of non-relativistic matter}
\label{sec:flow}
Here we present the formalism for describing non-relativistic matter flows
in cosmology and derive the matter density fluctuation $\dtm$ in the 
proper-time hypersurface.

\subsection{Spacetime metric}
\label{ssec:metric}
We first define the spacetime metric $g_{ab}$, on which our calculations rely.
The background universe is described by the usual FRW metric and 
small departures from the homogeneous and isotropic universe are captured by
metric perturbations
\beeq
\label{eq:abc}
\delta g_{00}=-2~\AA~, \quad
\delta g_{0\alpha}=-a~\BB_\alpha ~, \quad
\delta g_{\alpha\beta}=2~a^2\CC_{\alpha\beta}~,
\eneq
where the zeroth coordinate is the proper time~$t$ (not the conformal time), 
the scale factor is $a(t)$, and the perturbations 
$\BB_\alpha$ and $\CC_{\alpha\beta}$ are based on the three-metric
$\gbar_{\alpha\beta}$ in the background. The departures in the
metric are defined in a non-perturbative way,
and hence each variable can be perturbatively split at each order, e.g.,
\beeq
\AA=\AA^{(1)}+\AA^{(2)}+\AA^{(3)}+\cdots~.
\eneq
According to the generalized Helmholtz equation \cite{BARDE80,BARDE88}, 
we further decompose
the perturbation variables into scalar ($\beta$, $\varphi$, $\gamma$), 
transverse vector ($\VBB_\alpha$, $\VCC_\alpha$), and traceless transverse
tensor $(\TCC_{\alpha\beta}$) as
\beeq
\label{eq:svt}
\BB_\alpha=\beta_{,\alpha}+\VBB_\alpha~,\quad
\CC_{\alpha\beta}=\varphi~\gbar_{\alpha\beta}+\gamma_{,\alpha|\beta}+
\VCC_{(\alpha|\beta)}+\TCC_{\alpha\beta}~,
\eneq
where the round bracket is the symmetrization and
the comma and the vertical bar are the spatial derivative and
the covariant derivative with respect to $\gbar_{\alpha\beta}$, respectively.
It is noted that the decomposition is also 
independent of perturbation orders \cite{YOZA14}, and their spatial
indices make the separation of scalar, vector, and tensor apparent.

\subsection{ADM formalism}
\label{ssec:adm}
As we need to work on higher-order perturbations, it proves convenient
to work with the Arnowitt-Deser-Misner (ADM) formalism \cite{ADM,MTW}
and to derive fully nonlinear equations before we perform perturbative 
calculations. Its connection to the perturbed FRW metric 
is given in Appendix~\ref{app:metric}.

In the ADM formalism, the spacetime is split into an ordered
sequence of hypersurfaces labeled by a time coordinate~$t$, and
the intrinsic geometry of hypersurfaces is represented by the spatial metric
$h_{\alpha\beta}=g_{\alpha\beta}$. The proper-time $\Delta\tp$
between two hypersurfaces separated by $\Delta t$ is characterized by
the lapse function~$N$, and the shift vector $N^\alpha$ describes
the spatial coordinate change of a normal direction between the hypersurfaces:
\beeq
\Delta\tp=N\Delta t~,\qquad \Delta x^\alpha=N^\alpha\Delta t~.
\eneq
Therefore, the spacetime metric in the ADM formalism is described as
\beeq
\label{eq:ADM}
ds^2=g_{ab}dx^adx^b
=-N^2dt^2+\gamma_{\alpha\beta}(dx^\alpha+N^\alpha dt)(dx^\beta+N^\beta dt)~,
\eneq
where the individual metric components are
\beeq
g_{00}=-N^2+N^\alpha N_\alpha~, \quad 
g_{0\alpha}=N_\alpha=h_{\alpha\beta}N^\beta~, \quad
g_{\alpha\beta}=\gamma_{\alpha\beta}~,
\eneq
and their inverse components are
\beeq
g^{00}=-{1\over N^2}~, \quad
g^{0\alpha}={N^\alpha\over N^2} ~,\quad
g^{\alpha\beta} =\gamma^{\alpha\beta} - { N^\alpha N^\beta\over N^2}~.
\eneq

Given the 3+1 split in the ADM formalism, the local bending of spacelike 
hypersurfaces in spacetime is described by the extrinsic curvature
\beeq
K_{\alpha\beta}={1\over 2N} \left( N_{\alpha:\beta}+ N_{\beta:\alpha} 
-\gamma_{\alpha\beta,0} \right)=-N\Gamma^0_{\alpha\beta}~,
\eneq
where $\Gamma^a_{bc}$ is the Christoffel symbol based on $g_{ab}$ and
the colon is the covariant derivative with respect to $\gamma_{\alpha\beta}$.
The extrinsic curvature can be further split into the trace part
$K=\gamma^{\alpha\beta} K_{\alpha\beta}$ and the
traceless part~$\bar K_{\alpha\beta}$ as
\beeq
\bar K_{\alpha\beta}=K_{\alpha\beta}- {1\over 3} \gamma_{\alpha\beta} K~.
\eneq

\subsection{Different observers}
\label{ssec:obs}
In cosmology, many different observers can be defined in describing the
fluid and the metric quantities, although each observer may not
be related to real observation. Here we clarify the difference by providing
the exact definitions for later use, 
while keeping the terminology ``observers.''

In the ADM formalism, the normal observer is defined by the flow of the
normal direction of spatial hypersurfaces
\beeq
\label{eq:normal}
n_a=(-N,0)~,\qquad n^a=\left({1\over N},~ - {1\over N} N^\alpha\right)~.
\eneq
The normal observer is indeed the normal vector of hypersurfaces in 3+1
split, the flow of which is related to the extrinsic curvature
\beeq
K_{\alpha\beta} =-n_{\alpha;\beta}~,
\eneq
where the semicolon denotes the covariant derivative with respect to $g_{ab}$
in spacetime. The induced metric on the hypersurface is, therefore,
$\gamma_{\alpha\beta}=g_{\alpha\beta}+n_\alpha n_\beta=g_{\alpha\beta}$.
The normal observer, as defined in a given coordinate system, is a geometric
quantity, but is not necessarily related to any flow of matter.

In general, a four velocity vector $\UU^a$ can be defined to describe the
flow of any fluids in cosmology
\beeq
\label{eq:four}
\UU^0\equiv{1+\dUU}~,\qquad
\UU^\alpha\equiv{1\over a}\VV^\alpha~,
\eneq
where the perturbations $(\dUU,\VV^\alpha)$
are defined with respect to the case in a homogeneous
universe (hence based on $\gbar_{\alpha\beta}$) 
and they are subject to the time-like normalization
condition ($\UU^a\UU_a=-1$; similarly for the normal observer). 
If spatial velocity vector $\VV^\alpha$ is
the velocity of a fluid component, the observer described by
$u^\alpha$ moves together
with the fluid and is called the comoving observer. Here we will consider
the case in which the observer with $\UU^a$
always moves together with the fluid,
hence the comoving observer. However, 
two velocities can be different in principle, and the observer with $\UU^a$
may not be necessarily comoving with any fluids.

Another observer of interest
is the coordinate observer ($\VV^\alpha\equiv0$), 
whose motion is fixated at a given spatial coordinate (hence the name). 
Same as for the 
normal observer, the coordinate observer is not directly related to
any flow of matter. For later convenience, we define the covariant
spatial component of the four velocity vector
\beeq
\label{eq:ualpha}
\UU_\alpha= g_{\alpha b}\UU^b\equiv a\left(-\SVV_{,\alpha}+\VVV_\alpha\right)~,
\eneq
in terms of scalar $\SVV$ and vector $\VVV_\alpha$ components. 
The four velocity vector $\UU^a$ can be used to describe the normal vector 
$n^a$, if $\SVV=\VVV_\alpha=0$, which is called the comoving gauge condition
(see Sec.~\ref{sec:gauge}).

It is evident from the definition of various observers that the comoving
observer is physically relevant to the evolution of fluids in cosmology,
while the normal and the coordinate observers describe the geometry of a
given spacetime metric and coordinate system.

\subsection{Covariant decomposition and 
energy-momentum tensor}
\label{ssec:emt}
Any four velocity vector can be covariantly decomposed into 
physically well-defined quantities of flows described by
$\UU^a$ \cite{EHLER61,ELLIS71}
\beeq
\label{eq:cov}
\UU_{a;b}={1\over 3}\theta~\pr_{ab}+\sigma_{ab}+\omega_{ab}- a_a\UU_b~,
\eneq
where the expansion and the acceleration of the flow are 
$\theta={\UU^a}_{;a}$~, $a_a=\UU^b{\UU^a}_{;b}$~, the projection
tensor $\pr_{ab}=g_{ab}+\UU_a\UU_b$~, and 
the shear and the rotation of the flow are
$\sigma_{ab}=\UU_{(a;b)}+a_{(a}\UU_{b)}$~, and
$\omega_{ab}=\UU_{[a;b]}+a_{[a}\UU_{b]}$~.
A similar decomposition is possible for the normal observer~$n^a$, and these
covariant quantities represent the geometry of the hypersurface in spacetime:
\beeq
\theta=-K~,\quad \omega_{ab}=0~,\quad 
\sigma_{\alpha\beta}=-\bar K_{\alpha\beta}~.
\eneq

The energy-momentum tensor of fluids can be written in full generality
\cite{EHLER61,ELLIS71}
\beeq
\TT_{ab}=\rho~\UU_a\UU_b+ p~\pr_{ab}+2q_{(a}\UU_{b)}+\pi_{ab}~,
\eneq
where $\rho$ and $p$ are the energy density and isotropic pressure of the 
fluid, $q_a$ is the energy flux, and $\pi_{ab}$ is the anisotropic pressure. 
Those fluid quantities are ones measured by the observer described by $\UU^a$:
\bear
\rho&=&\TT_{ab}\UU^a\UU^b~,\qquad p={1\over3}\TT_{ab}\pr^{ab}~,\nonumber \\
q_a&=&-\TT_{cd}\UU^c\pr^d_a~,\quad \pi_{ab}=\TT_{cd}\pr^c_a\pr^d_b-p~\pr_{ab}~,
\enar
and hence they are frame-dependent (or observer dependent) \cite{NOHW04}. 
Therefore, it is most
convenient to use the fluid quantities measured by the comoving observer,
or the fluid quantities in the rest frame, i.e., $q_a=0$.

Here we will focus on the presureless medium of non-relativistic matter
with $p=\pi_{ab}=0$,
a good approximation to the late-time Universe on large scales, 
where baryons are effectively pressureless. Hence the energy-momentum
tensor simplifies as
\beeq
\TT_{ab}=\rho_m\UU_a\UU_b~.
\label{eq:tam}
\eneq
Furthermore, we will consider an irrotational fluid $\omega_{ab}=0$,
which dictates that the vector component of the four velocity should vanish
\beeq
\VVV_\alpha=0~.
\eneq

\subsection{Geodesic  motion of non-relativistic matter}
\label{ssec:path}
We are interested in the motion of non-relativistic matter
described by the energy-momentum tensor in Eq.~(\ref{eq:tam}). Without 
pressure, the non-relativistic matter
responds to the gravity only, following
the geodesic path, and the geodesic equation is $a_\alpha=0$:
\bear
\label{eq:geo}
a_\alpha&=&\dot\UU_\alpha \UU^0+\UU_{\alpha,\beta}\UU^\beta
-{\UU_0\UU^0\over N}\left[N_{,\alpha}-K_{\alpha\beta}N^\beta\right]\nonumber \\
&&
-\UU_\beta \UU^0\left[- {1\over N} N_{,\alpha} N^\beta- N K^\beta_\alpha 
+ N^\beta_{\;\;\; :\alpha} + {1\over N} N^\beta N^\delta
K_{\alpha\delta}\right]\nonumber \\
&&
+{\UU_0\UU^\beta\over N}K_{\alpha\beta}-\UU_\delta\UU^\beta
\left[{\Gamma^{(\gamma)}}^\delta_{\alpha\beta}+{1\over N} 
N^\delta K_{\alpha\beta}\right]~,
\enar
where ${\Gamma^{(\gamma)}}^\delta_{\alpha\beta}$ is the Christoffel symbol
based on $\gamma_{\alpha\beta}$. For the normal observer 
$\UU_\alpha=n_\alpha=0$, the geodesic equation greatly simplifies as
\beeq
\label{eq:aN}
a_\alpha={1\over N}N_{,\alpha}=0~.
\eneq

In Sec.~\ref{ssec:obs} we considered different observers with four velocity
vector $\UU^a$. For non-relativistic
matter flows, the path described by $\UU^a$ is timelike,
and the normalization condition ($-1=\UU^a\UU_a$) implies that the path can
be parametrized by the affine parameter~$\cc$ in proportion to the proper
time~$\tp$, i.e., $\UU^a=dx^a/d\cc$. Therefore, 
the path of the observers in spacetime can be obtained by
integrating their velocity vector over the affine parameter~$\cc$
\beeq
\label{eq:geoo}
x^a_\cc-x^a_{\cc_o}=(t_\cc-t_{\cc_o},~x^\alpha_\cc)=\int_{\cc_o}^\cc
d\cc'~\UU^a~,
\eneq
where we set the spatial coordinate $x^\alpha_{\cc_o}=0$ at $\cc_o$. 
To the zeroth order in perturbation,
the spatial position remains unchanged $\delta x^\alpha=0$, and 
the proper time elapsed along the fluid is related to the affine parameter 
\beeq
\Delta\tp=\bar t_\cc-\bar t_{\cc_o}=\cc-\cc_o~.
\eneq
In the presence of perturbations, the path of the observers drifts
away from the background relation
\beeq
x^a_\cc-x^a_{\cc_o}\equiv(\Delta\tp+\delta\tp,~\delta x^\alpha)~,
\eneq
and from Eq.~(\ref{eq:geoo})
the spacetime drifts $\Delta x^a=(\delta\tp,~\delta x^\alpha)$ are derived
to the second order in perturbations as
\bear
\label{eq:dev}
\delta\tp&=&\int_{\bar t_{\cc_o}}^{\bar t}d\bar t
\bigg[\dUU+\Delta x^a{\dUU}_{,a}\bigg]~,\\
\delta x^\alpha&=&
\int_{\bar\eta_{\cc_o}}^{\bar\eta}d\bar\eta\bigg[\VV^\alpha+a\Delta x^b
\left({\VV^\alpha\over a}\right)_{,b}\bigg]~,\nonumber
\enar
where $\eta$ is the conformal time and the overbar is used to indicate that
the integration is along the background path. Note that the spacetime
drifts $\Delta x^a$ in the integrand
should be evaluated at~$\cc$, not at the background.
However, to the second order in perturbation, it can be evaluated at the 
background.
While we focus on the geodesic motion of non-relativistic matter, the spacetime
drifts in Eq.~(\ref{eq:dev}) are valid for flows with non-vanishing
acceleration, as long as their path is timelike.

\begin{table*}
\caption{Gauge conditions considered in this paper}
\begin{ruledtabular}
\begin{tabular}{cccccc}
gauge choice & temporal gauge condition & spatial gauge condition 
& ADM variables & comoving observer & remaining gauge mode\\
\hline
\gI   & comoving $\SVV=0$ & $\gamma=\VCC_\alpha=0$ (C-gauge)
& $N=1$, $N_\alpha\neq0$
& normal & No \\
\gII  & comoving $\SVV=0$ & $\beta=\VBB_\alpha=0$ (B-gauge)
& $N=1$, $N_\alpha=0$ 
& normal, coordinate &Yes \\
\gIII & synchronous $\AA=0$ & $\beta=\VBB_\alpha=0$ (B-gauge)
& $N=1$, $N_\alpha=0$ & $\cdot$ &Yes \\
\end{tabular}
\end{ruledtabular}
\label{tab:gauge}
\end{table*}

\subsection{Matter fluctuation in proper-time hypersurface}
\label{ssec:prop}
Having related the coordinate time~$t$ of observers to their locally
measured proper-time~$\tp$,
we can construct a hypersurface of same proper-time of non-relativistic matter
and compute the matter density fluctuation $\dtm$
in the proper-time hypersurface.
Since the matter density at a given spacetime point can be split into the 
background and the fluctuation around it,
\beeq
\rho_m(x^a)=\bar\rho_m(t)\left[1+\delta_m(x^a)\right]=\bar\rho_m(\tp)
\left[1+\dtm\right]~,
\eneq
we derive $\dtm$ to the second order in perturbation:
\beeq
\dtm=\delta_m-3H\delta\tp(1+\delta_m)+{3\over2}(3H^2-\dot H)\delta\tp^2~,
\label{eq:dtm}
\eneq
where $H$ is the Hubble parameter.
It is noted that the expression is gauge-invariant at the linear order, 
as the time-slicing is fully specified. A proper-time
hypersurface of non-relativistic matter flows is physically well-defined,
corresponding to a complete choice of gauge condition. 
However, at the second order, the spatial gauge transformation affects
perturbations, and the
spacing of the hypersurface needs to be fully specified.

\section{Gauge choice}
\label{sec:gauge}
Here we describe gauge choices one can make in computing the matter density
fluctuation $\dtm$ in the proper-time hypersurface.
Since the proper-time hypersurface of non-relativistic matter is physically
well-defined, any gauge choice can be made to compute $\dtm$ in 
Eq.~(\ref{eq:dtm}). However, it would be preferable to make a gauge choice,
in which the coordinate time represents the proper time of non-relativistic
matter flows, i.e., $\delta\tp=0$. Meanwhile, unphysical gauge modes 
may remain in the solutions for certain choices of gauge conditions. For
instance, 
it is well-known that the synchronous gauge fails to completely fix gauge
freedom and has gauge modes to the linear order (e.g., \cite{MABE95}).
Here we consider three popular choices of gauge
conditions summarized in Table~\ref{tab:gauge} and discuss the geodesic
motion of the comoving observer and the matter density fluctuation in each
gauge choice. The second-order matter density fluctuations are derived
in Sec.~\ref{sec:non}, and their gauge issues and physical
interpretation are discussed in Sec.~\ref{sec:issue} and Sec.~\ref{sec:phy},
respectively.

\subsection{Gauge transformation}
\label{ssec:gt}
The principle of general covariance dictates that any coordinate system
can be used to describe physics in general relativity. However, since
the background quantities in cosmology depend only on the time coordinate
due to symmetry, a change in coordinate systems accompanies a change 
in the correspondence to the background, and perturbations in a given 
coordinate system  accordingly change \cite{BARDE80}. 

Given a coordinate 
transformation,\footnote{Here we use the conformal time~$\eta$ in considering 
a coordinate transformation, instead of the proper time~$t$. The relation 
for two different coordinate transformations can be readily derived as
\beeq
T_t=aT_\eta+{1\over2}a'T_\eta^2+\cdots~,
\eneq
where $T_t$ and $T_\eta$ are defined in relation to their coordinate 
transformations.}
\beeq
\tilde \eta=\eta+T~,\qquad \tilde x^\alpha=x^\alpha+\LL^\alpha~,
\label{eq:ct}
\eneq
the scalar and the
vector perturbations gauge transform to the linear order as
\bear
\label{eq:gt}
\tilde\AA&=&\AA-T'-\HH T~,\quad \tilde\beta=\beta-T+L'~,\quad
\tilde\varphi=\varphi-\HH T~, \nonumber \\
\tilde\gamma&=&\gamma-L~, \qquad 
\tilde\SVV=\SVV-T~,\qquad
\tilde \delta_m=\delta_m+3\HH T~,\nonumber \\
\tilde \chi&=&\chi-aT~,\quad \tilde\kappa=\kappa+\left(3\dot H+
{\Delta\over a^2}\right)aT~,\nonumber\\
\tilde\VBB_\alpha&=&\VBB_\alpha+L'_\alpha~, \quad
\tilde\VCC_\alpha=\VCC_\alpha-L_\alpha~,
\enar
where the prime is the derivative with respect to the conformal time,
the conformal Hubble parameter is $\HH=a'/a=aH$, and
we further decomposed the spatial transformation into scalar~$L$ and
vector $L^\alpha$ as
\beeq
\mathcal{L}^\alpha=L^{,\alpha}+L^\alpha~.
\eneq
For later reference, we defined $\chi=a(\beta+\gamma')$ and 
$\kappa=\delta K=3(H\AA-\dot\varphi)-\Delta\chi/a^2$. The spatial
vector $\VVV_\alpha$ and the tensor  $\TCC_{\alpha\beta}$ perturbations
are gauge-invariant at the linear order. It is noted that
the gauge-transformation relations in Eq.~(\ref{eq:gt}) are valid only
to the linear order, and we will consider second-order gauge-transformation
in Sec.~\ref{sec:issue}.

Gauge freedoms expressed in terms of $T$ and $\LL^\alpha$ need to be fully
removed by an appropriate choice of gauge conditions. Otherwise, perturbation
variables are  not uniquely defined, as illustrated in Eq.~(\ref{eq:gt}).
We use {\it temporal} and {\it spatial} gauge conditions to refer to the
gauge conditions fixing the temporal~$T$ and the spatial $\LL^\alpha$
gauge freedoms, respectively.

\subsection{Gauge choice \gI:
Temporal comoving and spatial C-gauge}
\label{ssec:I}
We consider the first gauge choice in Table~\ref{tab:gauge}, in which
the temporal gauge condition is set by $\SVV=0$ of non-relativistic matter
flows
and the spatial gauge condition is set by $\gamma=\VCC_\alpha=0$
(C-gauge).
With the irrotational condition of the fluid, the temporal gauge condition 
$\SVV=0$ implies the covariant spatial component of
the observer vanishes $\UU_\alpha=0$ in Eq.~(\ref{eq:ualpha}) 
and the energy-momentum tensor in Eq.~(\ref{eq:tam}) is 
\beeq
\TT_{ab}=N^2\rho_m\delta^0_a\delta^0_b~,\qquad \TT^0_\alpha=0~.
\eneq
Therefore, this gauge choice is often called the {\it comoving gauge}, 
as the rest-frame comoving
observer sees a vanishing energy flux $\TT^0_\alpha=0$. 
Furthermore, the four velocity vector of the comoving observer in this case
describes the normal observer $\UU^a=n^a$, which differs from 
the coordinate observer, though.

It is apparent in Eq.~(\ref{eq:gt}) that the temporal comoving gauge condition
sets $T=0$ and the spatial C-gauge condition sets $L^\alpha=L=0$,
completely eliminating the gauge freedom to the linear order. To the second
order in perturbation, we will choose $\SVV=0$ as our temporal gauge condition.
It was explicitly shown \cite{NOHW04,YOZA14} 
(see also Eqs.~[\ref{eq:st}] and~[\ref{eq:gt22}] in 
Sec.~\ref{sec:issue}) that the spatial gauge condition
$\gamma=\VCC_\alpha=0$ to the higher-order in perturbation completely
fixes the gauge freedom $T=L=L^\alpha=0$ if the linear-order gauge condition
that sets $T=0$ is chosen to the higher-order in perturbation.

From the geodesic condition in Eq.~(\ref{eq:aN}), the lapse function can be
set $N=N(t)=1$ (see Sec.~\ref{sec:issue} and 
Appendix~\ref{app:metric}). The metric perturbations and the ADM variables 
in this gauge choice are
\bear
\BB_\alpha&=&{1\over a}\chi_{,\alpha}+\pv_\alpha~,\quad
\CC_{\alpha\beta}=\varphi\gbar_{\alpha\beta}+\TCC_{\alpha\beta}~,\\
N_\alpha&=&-\chi_{,\alpha}-a\pv_\alpha~,\quad 
h_{\alpha\beta}=a^2\left[(1+2\varphi)\gbar_{\alpha\beta}+2\TCC_{\alpha\beta}
\right]~,\nonumber
\enar
where we defined the vector perturbation
$\pv_\alpha=\VBB_\alpha+\VCC_\alpha'$. 
To the linear order in perturbation
$\pv_\alpha$ is gauge-invariant and $\chi$ is spatially gauge-invariant,
according to Eq.~(\ref{eq:gt}).

In this gauge choice, the comoving normal observer is
\beeq
\UU^a=n^a=(1,~-N^\alpha)~,\qquad \dUU=0~,
\eneq
and the geodesic path of the observer is
\beeq
x^a_\cc-x^a_{\cc_o}=(\Delta\tp,\delta x^\alpha)~,\qquad \delta\tp=0~,
\eneq
where the time drift vanishes and the spatial drift is
\beeq
\label{eq:sdI}
\delta x^\alpha=-
\int_{\bar t_{\cc_o}}^{\bar t}d\bar t\bigg[N^\alpha+\delta x^\beta 
{N^\alpha}_{,\beta}\bigg]~.
\eneq
Over some proper time $\Delta\tp$ measured by the comoving observer in the
rest frame of non-relativistic matter, they drift away from the initial
spatial position, but the time coordinate of the observer
in this gauge condition is {\it synchronized} with the proper time.
The matter density fluctuation $\dtm$
in the proper-time hypersurface is, therefore,
\beeq
\dtm=\delta_m^\up{\gI}~~~\up{for~gauge~choice~\gI}.
\eneq

To the linear order in perturbation, the coordinate observer is identical
to the comoving normal observer, and the same conclusion for $\dtm$
can be drawn.
However, at higher order, $\dUU$ of the coordinate observer is non-vanishing
(hence $\delta\tp\neq0$), and the coordinate observer has to
accelerate ($a_\alpha\neq0$) to stay at the same spatial coordinates.
Despite the non-geodesic motion of gauge choice~\gI, as described by the
coordinate observer, the matter density fluctuation in gauge choice~\gI\
represents that of the proper-time hypersurface, because it is the
comoving observer of non-relativistic matter that is physically relevant
and whose proper time is synchronized with coordinate time.

\subsection{Gauge choice \gII: 
Temporal comoving and spatial B-gauge}
\label{ssec:II}
The second gauge choice in Table~\ref{tab:gauge}
is a variant of the comoving gauge, which is identical
in the temporal gauge condition $\SVV=0$ but differs only in the spatial 
gauge condition $\BB_\alpha=0$ ($\beta=\VBB_\alpha=0$; B-gauge). Therefore,
the metric perturbations and the ADM variables in this gauge choice are 
\bear
\BB_\alpha&=&N_\alpha=0~,\quad
 h_{\alpha\beta}=a^2(\gbar_{\alpha\beta}+2~\CC_{\alpha\beta})~,\\
\CC_{\alpha\beta}&=&\varphi~\gbar_{\alpha\beta}+\gamma_{,\alpha|\beta}+
\VCC_{(\alpha|\beta)}+\TCC_{\alpha\beta}~,\nonumber
\enar
where no further simplification is possible for $\CC_{\alpha\beta}$.

With the same temporal
gauge condition, gauge choice~\gII\ is again the comoving gauge 
$\TT^0_\alpha=0$, and the comoving observer coincides with the normal observer.
With vanishing
shift function $N_\alpha=0$, it is also the coordinate observer 
($\VV^\alpha=0$) in this case.
The geodesic condition in Eq.~(\ref{eq:aN}) implies that
$N=1$ and $\AA=0$ (see Sec.~\ref{sec:issue}), and hence 
gauge choice~\gII\ is often called the {\it comoving-synchronous gauge}.

However, as is apparent in Eq.~(\ref{eq:gt}), while the temporal gauge freedom
is removed $T=0$, the spatial gauge freedom in this case 
is constrained only to its derivative, i.e., $\LL'_\alpha=0$, implying
that even to the linear order in perturbation 
there remain spatial gauge modes $L=L(\xvec)$ and $L_\alpha=L_\alpha(\xvec)$,
such that $\gamma$ and $\VCC_\alpha$ are uniquely determined up to 
any time-independent, but 
{\it scale-dependent} functions. We discuss how the remaining gauge modes
affect the solutions in Sec.~\ref{sec:issue}.

The motion of the comoving
observer is simpler in this gauge choice --- the observer
four velocity vector and its path are
\beeq
\UU^a=n^a=(1,0)~,\qquad x^a_\cc-x^a_{\cc_o}=(\Delta\tp,0)~,
\eneq
and with vanishing time drift $\delta\tp=0$
the matter density fluctuation $\delta_m$
in this gauge choice again represents the matter density fluctuation $\dtm$
in the same proper-time hypersurface
\beeq
\dtm=\delta_m^\up{\gII}~~~\up{for~gauge~choice~\gII}.
\eneq
For this simplicity, gauge choice~\gII\ has been widely used in literature
for computing nonlinear equations (e.g., \cite{KASAI92}). We show
in Sec.~\ref{sec:issue} that the remaining gauge modes in gauge choice~\gII\
affect the matter density fluctuation, and hence it is incomplete. 
While we can project out the gauge mode in $\delta_m^\up{\gII}$, we show
that the matter density fluctuations in 
gauge choices~\gI\ and~\gII\ are different in Sec.~\ref{sec:non}
 and its physical interpretation is presented in Sec.~\ref{sec:phy}.

\subsection{Gauge choice \gIII: The synchronous gauge}
\label{ssec:III}
The third gauge choice in Table~\ref{tab:gauge} is the original
{\it synchronous gauge},
in which the temporal gauge condition is set by
$\AA=0$ and the spatial gauge 
condition is $\BB_\alpha=0$ ($\beta=\VBB_\alpha=0$; B-gauge).
This gauge condition implies that the ADM variables are $N=1$ and $N_\alpha=0$.
Similarly in gauge choice~\gII,
the normal observer is the coordinate observer. However, the comoving observer
in this case is
\beeq
\label{eq:vvsyn}
\UU^a=\left(1+{1\over2}\VV^\alpha\VV_\alpha~,~{1\over a}\VV^\alpha\right)~,
\eneq
different from the normal observer. It is apparent that the geodesic motion
of the comoving observer is not synchronous with the coordinate time
($\delta\tp\neq0$) as $\dUU\neq0$,
and the matter density fluctuation $\delta_m$ in this gauge choice
differs from $\dtm$ of the proper-time hypersurface, unless
the spatial velocity vector vanishes $\VV^\alpha=0$. Nevertheless, 
Eq.~(\ref{eq:dtm}) can be used to compute $\dtm$ in gauge choice~\gIII,
despite $\delta\tp\neq0$ (hence $\dtm\neq\delta_m$).

Moreover, it is well-known that the synchronous gauge fails to 
fix the gauge freedom. To the linear order in perturbation, the gauge
transformation in Eq.~(\ref{eq:gt}) only constrains the gauge freedom as
\beeq
\label{eq:gtIII}
T={c_1(\xvec)\over a}~,\quad L=c_1(\xvec)\int^t{dt\over a^2}+c_2(\xvec)~,
\quad L_\alpha=L_\alpha(\xvec)~,
\eneq
where $c_i(\xvec)$ is a time-independent 
but scale-dependent function. Therefore,
the all the perturbation variables other than $\AA=\BB_\alpha=0$ have
remaining unphysical gauge modes, even at the linear order.

However, the geodesic condition in Eq.~(\ref{eq:geo}) yields 
that the spatial velocity vector decays in time
\beeq
\VV^\alpha\propto {1\over a}~,
\eneq
which suggests that by imposing the initial
condition $\VV^\alpha=0$ at some early time (e.g., see \cite{MABE95,SEZA96}),
the spatial velocity vanishes
all the time, and the comoving observer in Eq.~(\ref{eq:vvsyn})
becomes the coordinate observer (and the normal observer).
Indeed, this initial condition makes gauge
choice~\gIII\ identical to  gauge choice~\gII, as the covariant
spatial component also vanishes $\SVV=0$ (see Appendix~\ref{app:sync}).
Unless the comoving gauge condition $\SVV=0$ is imposed, 
gauge choice~\gIII\ is complicated and plagued with unphysical gauge modes 
beyond the linear order in perturbation. Hereafter, we assume the specific
initial condition is adopted
for gauge choice~\gIII\ 
(hence identical to gauge choice~\gII), and further discussion of gauge
choice~\gIII\ will be referred to gauge choice~\gII.

\section{Nonlinear evolution equations of the matter density fluctuation}
\label{sec:non}
Here we derive the nonlinear equation, governing the irrotational presureless
fluid \cite{HWNO05b,HWNO06b,HWNO07b,UGWA14} and obtain their solutions
for the gauge choices in Table~\ref{tab:gauge}. In both gauge choices~\gI\
and~\gII, the matter density fluctuations $\delta_m$ represent the
matter density fluctuation 
$\dtm$ of the proper-time hypersurface. However, we show that the 
second-order solutions $\delta_m$ in those gauge choices
are different from each other.

Using the covariant decomposition
in Eq.~(\ref{eq:cov}), the conservation of the energy-momentum tensor
in Eq.~(\ref{eq:tam}) yields that the irrotational presureless fluid
should follow the geodesic path
and the energy density is conserved along the geodesic motion:
\beeq
\label{eq:econ}
a_a=0~,\qquad {d\over d\cc}\rho+\rho~\theta=0~,
\eneq
where the derivative with respect to the affine parameter is 
$d/d\cc=u^b\partial_b$. It is noted that the geodesic condition used
in Sec.~\ref{sec:gauge} is the consequence of the energy-momentum conservation.
The evolution of the expansion~$\theta$ 
along the flow is described by the Raychaudhuri
equation \cite{RAYCH55}, and it simplifies for the irrotational pressureless 
medium as
\beeq
{d\over d\cc}\theta+{1\over3}\theta^2+\sigma_{ab}\sigma^{ab}
+R_{ab}\UU^a\UU^b=0~,
\eneq
where the Ricci tensor $R_{ab}$ can be further related to the energy-momentum
tensor $\TT_{ab}$ by using the Einstein equation
\beeq
R_{ab}\UU^a\UU^b=4\pi G\rho_m-\Lambda~.
\eneq

These nonlinear equations are sufficient to describe the evolution of the
irrotational pressureless fluid, and they can be readily solved by splitting
into the background and the perturbation. The background equations
are
\bear
0&=&\dot{\bar\rho}_m+3H\bar\rho_m~, \\
0&=&3(\dot H+H^2)+4\pi G\bar\rho_m-\Lambda~,\nonumber
\enar
and the nonlinear perturbation equations are
\bear
\label{eq:pert}
&&\dot\delta_m-\kappa=N^\alpha\delta_{m,\alpha}+\delta_m\kappa~,\\
&&\dot\kappa+2H\kappa-4\pi G\bar\rho_m\delta_m=
N^\alpha\kappa_{,\alpha}+{1\over3}\kappa^2+\sigma^{ab}\sigma_{ab}~,\nonumber
\enar
where the expansion of the normal observer is related to the perturbation
$\kappa=\delta K$ of the extrinsic curvature~$K$ as
\beeq
\theta=-K=3H-\kappa~.
\eneq
Combining the two equations, the differential equation for the evolution
of non-relativistic matter can be derived as
\bear
\label{eq:comb}
&&\ddot\delta_m+2H\dot\delta_m-4 \pi G \bar\rho_m\delta_m\\
&&
={1\over a^2}\left[a^2\left(N^\alpha\delta_{m,\alpha}+\delta_m\kappa
\right)\right]^\cdot+N^\alpha\kappa_{,\alpha}+{1\over3}\kappa^2+\sigma^{ab}
\sigma_{ab}~.\nonumber
\enar

These equations are derived by assuming the temporal comoving gauge
($\SVV=0$) with the normal observer $(\UU^a=n^a)$, while the spatial gauge
condition is left unspecified. Therefore, they apply to both gauge choices~\gI\
and~\gII\ in Table~\ref{tab:gauge}.

\subsection{Gauge choice \gI: 
Temporal comoving and spatial C-gauge}
\label{ssec:solI}
In gauge choice~\gI, the shift function $N^\alpha$ is given in 
Sec.~\ref{ssec:I}, and the source terms in Eq.~(\ref{eq:pert}) are
computed in Appendix~\ref{app:metric}. Therefore, the nonlinear 
evolution equations for the matter density fluctuation $\delta_m$ and the 
expansion perturbation~$\kappa$ can be written in terms of metric to the
second order in perturbation as
\begin{widetext}
\bear
\label{eq:mastI}
&&\dot\delta_m-\kappa=-{1\over a^2}\chi^{,\alpha}
\delta_{m,\alpha}-{1\over a}\pv^\alpha\delta_{m,\alpha}+\delta_m\kappa~,\\
&&\dot \kappa+ 2 H \kappa-4 \pi G \bar\rho_m \delta_m=
-{1\over a^2}\chi^{,\alpha}\kappa_{,\alpha}
-{1\over a}\pv^\alpha\kappa_{,\alpha}+
\left({1\over a^2}\chi_{,\alpha|\beta}+{1\over a}\pv_{\alpha|\beta}
+\dot\TCC_{\alpha\beta}\right)
\left({1\over a^2}\chi^{,\alpha|\beta}+{1\over a}\pv^{\alpha|\beta}
+\dot\TCC^{\alpha\beta}\right)~,\nonumber
\enar
\end{widetext}
where we used $\dot\varphi=0$ at the linear order in computing the quadratic
source
terms. These coupled evolution equations are sourced not only by the scalar
contributions, but also by the vector $\pv_\alpha$ and the tensor 
$\TCC_{\alpha\beta}$ contributions. While the evolution equations for the
vector and tensor contributions can be supplemented, we simplify the 
nonlinear evolution equations~(\ref{eq:mastI}) by neglecting the
{\it linear-order}
vector and tensor contributions, as we are interested in the evolution of
non-relativistic matter in the late time. Furthermore, since the
expansion perturbation is $\kappa=-\nabla^2\chi/a^2$ at the linear order (see 
Appendix~\ref{app:metric}), the nonlinear evolution
equations become a closed system of two differential equations for
$\delta_m$ and~$\kappa$.

To the second order in perturbation, we define a velocity vector~$\vvec$ in
relation to the expansion perturbation as
\beeq
\label{eq:newtV}
\kappa\equiv-{1\over a}\nabla\cdot\vvec~,
\eneq
and the velocity vector is curl-free to the linear order:
\beeq
\vvec^{(1)}={1\over a}\nabla\chi^{(1)}~.
\eneq
In terms of the velocity vector, the nonlinear evolution equations for
the matter density fluctuation and the expansion perturbation 
in gauge choice~\gI\ become identical
to the Newtonian equations for a presureless medium \cite{HWNO99,HWNO05}:
\bear
\label{eq:newtDI}
&&\dot\delta_m+{1\over a}\nabla\cdot\vvec=-{1\over a}\nabla\cdot(\delta_m
\vvec)~,\\
&&\nabla\cdot\dot\vvec+H\nabla\cdot\vvec+4\pi Ga\bar\rho_m\delta_m=
-{1\over a}\nabla\cdot\left[(\vvec\cdot\nabla)\vvec\right]
~,\nonumber
\enar
if we identify the matter density fluctuation~$\delta_m$ and the expansion
perturbation~$\kappa$ in gauge choice~\gI\ as the Newtonian matter density
fluctuation and the Newtonian expansion $\Theta=\nabla\cdot\vvec$ 
as in Eq.~(\ref{eq:newtV}). With this identification,
the master equation for the matter density fluctuation in 
Eq.~(\ref{eq:comb}) can be rephrased as
\bear
\label{eq:combI}
&&\ddot\delta_m+2H\dot\delta_m-4 \pi G \bar\rho_m\delta_m \\
&&\hspace{20pt}
=-{1\over a^2}\left[a\nabla\cdot(\delta_m\vvec)\right]^\cdot
+{1\over a^2}\nabla\cdot\left[(\vvec\cdot\nabla)\vvec\right]~.\nonumber
\enar

The correspondence between the Newtonian dynamics 
and the relativistic dynamics  in gauge choice~\gI\
is valid only to the second order in perturbation in the absence of 
{\it linear-order vector} and {\it tensor components}
in a universe with irrotational pressureless medium,
while it is noted that the Newtonian equations as in Eq.~(\ref{eq:newtDI}) are
fully nonlinear (see, e.g., \cite{BECOET02}), 
valid to all orders in perturbation. Beyond the second order, however, no exact
correspondence is possible due to relativistic corrections
(see, e.g., \cite{HWNO05}).
Second-order vectors and tensors are naturally generated by scalar
contributions, and they affect the third-order scalar contributions.
However, in the absence of the linear-order vector or tensor components,
scalar-generated second-order vector or tensor components are decoupled
from the nonlinear evolution equations~(\ref{eq:mastI}).

\subsection{Gauge choice \gII: 
Temporal comoving and spatial B-gauge}
\label{ssec:solII}
In gauge choice~\gII, the metric perturbations are present only in the spatial
metric $h_{\alpha\beta}$, and
the nonlinear evolution equations~(\ref{eq:pert}) and~(\ref{eq:comb})
are simpler due to the absence of the shift function $N^\alpha$. In terms of
metric perturbations, the first two source terms in the right-hand side of
Eqs.~(\ref{eq:mastI}) are absent in gauge choice~\gII, but its generic 
structure of the coupled differential equations
remains unchanged --- the evolution equations are closed, only if linear-order
vector and tensor contributions vanish.

Under the same assumption that {\it no} 
vector and tensor contributions are present
at the linear order, we express the evolution equations
in terms of the Newtonian velocity vector defined in Eq.~(\ref{eq:newtV}) as
\bear
\label{eq:newtDII}
&&\dot\delta_m+{1\over a}\nabla\cdot\vvec=-{1\over a}\delta_m\nabla\cdot\vvec
~,\\
&&\nabla\cdot\dot\vvec+H\nabla\cdot\vvec+4\pi Ga\bar\rho_m\delta_m
=-{1\over a}\nabla\cdot\left[(\vvec\cdot\nabla)\vvec\right]\nonumber \\
&&\hspace{30pt}
+{1\over a}\vvec\cdot\nabla(\nabla\cdot\vvec)~,\nonumber
\enar
and the master equation for the matter density fluctuation in 
Eq.~(\ref{eq:comb}) becomes
\bear
\label{eq:combII}
&&\ddot\delta_m+2H\dot\delta_m-4 \pi G \bar\rho_m\delta_m \\
&&
=-{1\over a^2}\left[a\delta_m\nabla\cdot\vvec\right]^\cdot+
{1\over a^2}\nabla\cdot\left[(\vvec\cdot\nabla)\vvec\right]
-{1\over a^2}\vvec\cdot\nabla(\nabla\cdot\vvec)~. \nonumber
\enar
As in gauge choice~\gI, there exists a Newtonian correspondence in 
the evolution equations in gauge choice~\gII\ \cite{HWNO06b}. As we showed
in Sec.~\ref{sec:gauge}, the spatial coordinates of
non-relativistic matter remain
unchanged all the time (the spatial drift $\delta x^\alpha=0$), such that
the coordinate system is tracking the particle motion, i.e., if we identify
the time derivative in Eq.~(\ref{eq:newtDII}) as the Lagrangian derivative
in Newtonian dynamics
\beeq
{\partial \over \partial t}~\RA~{D\over Dt}={\partial\over \partial t}+
\left(\vvec\cdot\nabla\right)~,
\eneq
the evolution equations can be recasted as those in the Newtonian Lagrangian
frame \cite{HWNO06b,HWNOET14,BRHIET14}, describing the same system as
in gauge choice~\gI. 
However, this correspondence is again valid only to the
second order in perturbation, in the absence of linear-order
vector or tensor contributions. At the third order in perturbation,
pure relativistic corrections appear \cite{HWNOET14}.

\subsection{Nonlinear solutions for matter density and expansion
perturbation}
\label{ssec:sol}
Having set up the closed coupled differential equations, 
we derive the solutions for the matter density fluctuation and the expansion
perturbation with the gauge choices in 
Table~\ref{tab:gauge}. As the evolution equations are rephrased in a way
similar to the Newtonian dynamics, we can simply follow the standard 
perturbative approach to solving the differential equations (e.g., 
\cite{BECOET02}).

The matter density fluctuation and the divergence of the velocity vector
(or the expansion) are expanded in Fourier space as
\bear
\delta_m(\kvec,t) &=& \delta_m^{(1)}+\delta_m^{(2)}+\cdots ~,\\
\Theta(\kvec,t)&=&\left[\nabla\cdot\vvec\right](\kvec,t)
=\Theta^{(1)}+\Theta^{(2)}+\cdots~, \nonumber
\enar
and the linear-order solutions takes the usual form
\beeq
\label{eq:lin}
\delta_m^{(1)}(\kvec,t)=D(t)\delta(\kvec)~,\quad 
\Theta^{(1)}(\kvec,t)=-\HH D(t)f\delta(\kvec)~,
\eneq
where $\delta(\kvec)$ is the matter density fluctuation at the initial time,
the growth factor $D(t)$ is normalized at the initial time, and 
$f=d\ln D/d\ln a$ is the logarithmic growth rate.\footnote{Even in the 
Newtonian dynamics, it is difficult to obtain an exact analytic
solution for the matter density fluctuation, unless the Universe is a
Einstein-de~Sitter (EdS) universe with maximum symmetry. However, it is
well-known that good 
approximate solutions are available in analytic form, when the growth 
factor~$D$ and the logarithmic growth rate~$f$ 
are used in the perturbative approach.} 
The master equations~(\ref{eq:combI}) and~(\ref{eq:combII})
for the matter density fluctuation in gauge choices~\gI\ and~\gII\ 
are identical to the linear order, accommodating
the same linear-order solution, as in Eq.~(\ref{eq:lin}).

The second-order solution can be derived  by using the standard
convolution forms in Fourier space as
\bear
\delta_m^{(2)}(\kvec,t)
&=&D^2(t)\int{d^3\qvec\over(2\pi)^3}
F_2(\qvec,\kvec-\qvec)\delta(\qvec)\delta(\kvec-\qvec) ~,\\
\Theta^{(2)}(\kvec,t) &= & -\HH fD^2(t)\int{d^3\qvec\over(2\pi)^3}
G_2(\qvec,\kvec-\qvec)\delta(\qvec)\delta(\kvec-\qvec) ~.\nonumber
\enar
For gauge choice~\gI,
the perturbation kernels for the matter density fluctuation and the
velocity divergence are
\bear
\label{eq:kernI}
F_2^\up{\gI}(\qvec_1,\qvec_2)
&=&{5\over7}+{2\over7}\left({\qvec_1\cdot\qvec_2\over q_1q_2}
\right)^2+{\qvec_1\cdot\qvec_2\over2q_1q_2}\left({q_1\over q_2}+{q_2\over q_1}
\right)~,~~~~~~\\
G_2^\up{\gI}(\qvec_1,\qvec_2)
&=&{3\over7}+{4\over7}\left({\qvec_1\cdot\qvec_2\over q_1q_2}
\right)^2+{\qvec_1\cdot\qvec_2\over2q_1q_2}\left({q_1\over q_2}+{q_2\over q_1}
\right)~.\nonumber
\enar
Since the governing equations~(\ref{eq:newtDI}) coincide with the Newtonian
dynamics, the perturbation kernels are naturally
identical to those in the Newtonian
perturbation theory. For gauge choice~\gII,
as the source terms in Eq.~(\ref{eq:newtDII}) are different from the standard
Eulerian perturbation theory, the perturbation kernels are also different
(see \cite{BIGOJE14,HWNOET14,RAMPF14})
\bear
\label{eq:kernII}
F_2^\up{\gII}(\qvec_1,\qvec_2)
&=&{5\over7}+{2\over7}\left({\qvec_1\cdot\qvec_2\over q_1q_2}\right)^2~,\\
G_2^\up{\gII}(\qvec_1,\qvec_2)
&=&{3\over7}+{4\over7}
\left({\qvec_1\cdot\qvec_2\over q_1q_2}\right)^2~.\nonumber
\enar
Compared to the perturbation kernels in Eqs.~(\ref{eq:kernI}) for
gauge choice~\gI, the dipole terms are absent in Eqs.~(\ref{eq:kernII})
for gauge choice~\gII. The matter density fluctuations in the gauge choices
in Table~\ref{tab:gauge} are, therefore, different at the second-order.
We discuss the difference in
their physical interpretation in Sec.~\ref{sec:phy}.

As is the case in the standard perturbation theory \cite{GOGRET86,BECOET02}
the recurrence relation can be derived for the perturbation kernels 
$(F_n,G_n)$
at higher order in gauge choice~\gI\ (e.g., \cite{JEGOET11,BIGOJE14}).
However, in the relativistic dynamics, the nonlinear equations~(\ref{eq:pert})
are closed to the second order
only when the linear-order vector and tensor contributions are neglected.
Furthermore, 
the shear amplitude $\sigma^{ab}\sigma_{ab}$ in the source term has
additional contributions from $\delta_m$ and $\vvec$ at orders beyond the
second order \cite{HWNO05}, invalidating the use of the recurrence relation
at orders higher than that of the shear amplitude. More importantly, beyond the
second order in perturbation, scalar-generated vector and tensor contributions
may cause systematic errors in the higher-order calculations.

\section{Gauge issues in the solution}
\label{sec:issue}
Gauge issue is a flaw in theory, and it has to be removed before any 
unique prediction in theory can be made in comparison to physical quantities.
As we discussed in Sec.~\ref{sec:gauge}, there exist remaining gauge modes
in gauge choice~\gII. We show that despite their presence, the matter density
fluctuation in gauge choice~\gII\ can be made independent
to the second order in perturbation.

\subsection{Linear-order gauge-transformation}
\label{ssec:lgt}
Both gauge choices~\gI~and~\gII\ in Table~\ref{tab:gauge}
take as the temporal gauge condition
the comoving gauge $\TT^0_\alpha=0$, which to the linear order
in perturbation imposes the vector component of the four velocity vanishes
$\tilde\VVV_{\alpha}=\VVV_{\alpha}=0$ and the scalar component satisfies
\beeq
\tilde\SVV_{,\alpha}=\SVV_{,\alpha}+T_{,\alpha}=0~.
\eneq
Furthermore, the momentum constraint of the pressureless fluid indicates that
the flow follows the geodesic in Eq.~(\ref{eq:aN})
\beeq
0=N_{,\alpha}=\AA_{,\alpha}~,
\eneq
where we used the ADM relation to the metric perturbations in 
Appendix~\ref{app:metric}. 
The temporal gauge freedom is indeed constrained by the comoving gauge 
condition, only to be a 
scale-independent, but time-dependent function $T=T(t)$, which can be used
to set $\AA=0$ by specifying $T=c/a(t)$, where $c$ is some (unspecified) 
constant. This constant is further removed ($c=0$, hence $T=0$)
by the conservation of the curvature perturbation $\dot\varphi=0$.

Following these series of gauge transformations, both gauge choices have
vanishing metric perturbation $\AA=0$ in time coordinates, and the time lapse
of the ADM variable is $N=1$. However, this is a deliberate gauge choice,
not {\it automatically} imposed by the comoving gauge condition.
Two gauge choices in Table~\ref{tab:gauge} differ in the spatial gauge 
condition, but the remaining spatial gauge mode in gauge choice~\gII\
has no impact on the matter density fluctuation~$\delta_m$ and the 
expansion perturbation~$\kappa$ as shown in Eq.~(\ref{eq:gt}). 
This is further borne out by the equivalence of the nonlinear evolution
equations~(\ref{eq:pert}) at the linear order.

\subsection{Second-order gauge-transformation}
\label{ssec:ngt}
To the second order in perturbation, the temporal gauge freedom can be 
completely removed ($T=0$) for both gauge choices in Table~\ref{tab:gauge}
in a similar way to the linear-order case
by setting $N=1$ from the geodesic condition, in addition to the comoving
gauge condition $\SVV=\VVV_\alpha=0$. This implies that the metric perturbation
in the time component vanishes $\AA=0$ for gauge choice~\gII, while only
the combination vanishes for gauge choice~\gI
\beeq
\AA+{1\over2}\BB^\alpha\BB_\alpha=0~,
\eneq
where we used $\AA^{(1)}=0$ (see Appendix~\ref{app:metric}). However,
the above combination can be set to be a non-vanishing, but scale-independent
function. 

Regarding the spatial gauge condition, gauge choice~\gI\ removes 
$L^{(1)}=L^{(1)}_\alpha=0$ in Eq.~(\ref{eq:gt}), and the spatial metric
perturbation in this case transforms to the second order in perturbation as 
\cite{NOHW04,YOZA14}
\beeq
\label{eq:st}
\tilde\CC^{(2)}_{\alpha\beta}=\CC^{(2)}_{\alpha\beta}
-\LL^{(2)}_{(\alpha|\beta)}~.
\eneq
Therefore, the decomposed perturbations transform as
\beeq
\label{eq:gt22}
\tilde\gamma^{(2)}=\gamma^{(2)}-L^{(2)}~,\quad
\tilde\VCC^{(2)}_\alpha=\VCC^{(2)}_\alpha-L^{(2)}_\alpha~,
\eneq
and the spatial gauge freedom can be completely removed in gauge choice~\gI\
by imposing $\gamma=\VCC_\alpha=0$ to the second order in perturbation.

By contrast, gauge choice~\gII\ sets $L^{(1)\prime}=L_\alpha^{(1)\prime}=0$
in Eq.~(\ref{eq:gt}), and the off-diagonal metric perturbation in this case
transforms as
\beeq
\tilde\BB_\alpha^{(2)}=\BB_\alpha^{(2)}+\LL_\alpha^{(2)\prime}
-\BB_{\alpha|\beta}\LL^\beta-\BB_\beta{\LL^\beta}_{|\alpha}~.
\eneq
Therefore, the spatial gauge freedom is constrained in gauge choice~\gII\
as $\LL_\alpha^{(2)\prime}=0$, and the residual spatial gauge mode remains
as
\beeq
L=L^{(1,2)}(\xvec)~,\quad L_\alpha=L^{(1,2)}_\alpha(\xvec)~.
\eneq

In gauge choice~\gI, no gauge freedom remains, and the solutions of the 
nonlinear Eqs.~(\ref{eq:pert}) and hence Eqs.~(\ref{eq:newtDI}) are uniquely
determined. In gauge choice~\gII, the remaining spatial gauge freedom can
affect the solution, as the matter density fluctuation and the expansion
transform to the second order in perturbation as \cite{HWNO06b}
\bear
\label{eq:gm}
\tilde\delta^\up{\gII}_m(\xvec,t)&=&\delta^\up{\gII}_m(\xvec,t)
-\nabla\delta^\up{\gII}_m\cdot\LL^\alpha(\xvec)~,\\
\tilde\kappa^\up{\gII}(\xvec,t)&=&\kappa^\up{\gII}(\xvec,t)
-\nabla\kappa^\up{\gII}\cdot\LL^\alpha(\xvec)~.\nonumber
\enar
It is now apparent that the solutions $\delta_m$ and $\kappa$ in gauge 
choice~\gII\ are {\it not} uniquely determined due to the arbitrary 
scale-dependent function~$\LL_\alpha(\xvec)$. However, it appears that
the solutions in Eq.~(\ref{eq:kernII}) are uniquely determined in gauge
choice~\gII, despite the presence of gauge modes.
Since the time-independent gauge modes in Eq.~(\ref{eq:gm}) are 
multiplied by the linear-order solutions, it vanishes in the left-hand side 
of Eqs.~(\ref{eq:newtDII}) and~(\ref{eq:combII}).
In other words, the solutions in Eq.~(\ref{eq:kernII}) are obtained by
projecting out the remaining gauge modes in Eq.~(\ref{eq:gm}).

\section{Physical interpretation of the solutions}
\label{sec:phy}
The matter density fluctuations $\delta_m^\up{\gI,\gII}$
in two gauge choices represent the matter density fluctuation $\dtm$ in the
same proper-time hypersurface, and they both lack any gauge issues.
At first glance, this conclusion appears odd, because
the proper-time hypersurface of non-relativistic matter is physically
well-defined and unique, yet solutions in two gauge choices differ as shown in
Eqs.~(\ref{eq:kernI}) and~(\ref{eq:kernII}). The power spectrum and the 
bispectrum of the matter density fluctuations in these gauge choices are
computed in \cite{JEGOET11,BIGOJE14}, showing the clear difference in two
gauge choices. As we discussed in Sec.~\ref{sec:non}, the relativistic
dynamics in gauge choices~\gI\ and~\gII\ are identical to the Eulerian
and the Lagrangian Newtonian dynamics, respectively. However, we again
emphasize that these correspondences are valid in the absence of linear-order
vector or tensor, and pure relativistic corrections appear beyond the
second-order in perturbations.

At the linear order, the gauge choices in Table~\ref{tab:gauge} are identical,
and the matter density fluctuations are also equivalent
$\delta_m^\up{\gI}=\delta_m^\up{\gII}$ 
(and $\kappa^\up{\gI}=\kappa^\up{\gII}$), which can be obtained from the
Boltzmann codes such as {\footnotesize CMBFAST} \cite{SEZA96},
{\footnotesize CAMB} \cite{LECHLA00}, and {\footnotesize CLASS} \cite{CLASS}.
The difference arises beginning at the second order in perturbation, and 
the critical difference can be found in the large-scale limit of their
kernels:
\bear
\lim_{\kvec\RA0}F_2^\up{\gI}(\qvec,\kvec-\qvec)&=&{3-5\mu^2\over7}{k^2\over 
q^2}+\OO(k^3)~,\\
\lim_{\kvec\RA0}F_2^\up{\gII}(\qvec,\kvec-\qvec)&=&1+{2(\mu^2-1)\over7}
{k^2\over q^2}+\OO(k^3)~,\nonumber
\enar
where $\mu=\kvec\cdot\qvec/kq$.
The kernel for gauge choice~\gI\ vanishes as $k^2$ in the large-scale
limit, while the kernel for gauge choice~\gII\ becomes unity.

It is argued \cite{PEEBL80} that any nonlinear correction to the initial
density field has to scale as wavenumber with power no less than two
in the large-scale limit, since gravity respects the mass and the momentum
conservation. The matter density fluctuation $\delta_m^\up{\gII}$ in 
gauge choice~\gII\ in this respect violates the mass conservation, which
essentially is due to the absence of the dipole term in Eq.~(\ref{eq:kernII}).
This can be further elaborated by considering the ensemble average of the
matter density fluctuations:
\beeq
\label{eq:sec}
\left\langle\delta^{(2)}_m(t,\xvec)\right\rangle=
\int{d^3\qvec\over(2\pi)^3}~P(t,\qvec)F_2(\qvec,-\qvec)=
\bigg\{\begin{array}{cc}0&\up{for~{\gI}}\\\sigma^2_m~&\up{for~\up{\gII}}
\end{array}~,
\eneq
where $P(t,\qvec)$ is the linear matter power spectrum and 
$\sigma_m^2=\langle\delta_m^2\rangle$ is the unsmoothed rms fluctuation.

To the second order, the matter density fluctuation in gauge 
choice~\gI, therefore, properly represents the ``mean'' matter 
density and the fluctuation around the mean:
\beeq
\label{eq:mean}
\rho_m(t,\xvec)=\bar\rho_m(t)\left(1+\delta_m^\up{\gI}\right)~,\quad
\langle\rho_m(t,\xvec)\rangle=\bar\rho_m(t)~,
\eneq
and it is noted that the mean matter density $\bar\rho_m$
is based on the coordinate time due to symmetry and its equality to the
ensemble average is {\it not}  by the definition. By contrast, the matter
density fluctuation in gauge choice~\gII\ has non-vanishing mean
\beeq
\label{eq:mean2}
\rho_m(t,\xvec)=\bar\rho_m(t)\left(1+\delta_m^\up{\gII}\right)~,\quad
\langle\rho_m(t,\xvec)\rangle=\bar\rho_m(t)\left(1+\sigma_m^2\right)~,
\eneq
as is derived in Eq.~(\ref{eq:sec}). 
The local observer sitting at the flow of non-relativistic matter has no
way to obtain the ``mean'' matter density $\langle\rho_m\rangle$ averaged
over the proper-time hypersurface. However, the mean matter density
$\bar\rho_m(\tp)$ in the homogeneous universe can be estimated by using
the proper time~$\tp$ of the observer. Therefore, the correct matter density
fluctuation $\dtm$ in the proper-time hypersurface is represented by the
matter density fluctuation $\delta_m^\up{\gI}$ in gauge choice~\gI.

However, it remains still puzzling that both gauge choices as shown in 
Sec.~\ref{sec:gauge} represent the physically well-defined hypersurface of 
non-relativistic matter flow, yet the matter density fluctuations differ
without any gauge issues present.
The resolution can be found by considering a transformation from gauge 
choice~\gII\ to gauge choice~\gI.
Both gauge choices share the common time coordinates $t^\up{\gI}=t^\up{\gII}$
($T=0$), but differ only in the spatial coordinates
\beeq
\label{eq:lpt}
\xvec^\up{\gI}=\xvec^\up{\gII}+\LL^\alpha(t,\xvec)~.
\eneq
According to Eq.~(\ref{eq:gt}), we derive the linear-order gauge transformation
as
\beeq
L=\gamma^\up{\gII}~,\quad L'={1\over a}\chi~,\qquad
L_\alpha=\VCC_\alpha^\up{\gII}~,\quad L_\alpha'=\pv_\alpha~,
\eneq
where $\pv_\alpha$ is gauge invariant and $\chi$ is spatially invariant
to the linear order. Integrating over time, the spatial transformation is
obtained as
\beeq
\label{eq:spt}
\LL^\alpha=\int^t {dt\over a}\left({1\over a}\nabla\chi
+\pv^\alpha\right)+\nabla\gamma^\up{\gII}(\xvec)+\VCC_\alpha^\up{\gII}(\xvec)~,
\eneq
where the integral term represents the time-dependent physical modes.
The remaining time-independent but scale-dependent metric perturbations 
represent the remaining gauge freedom in gauge choice~\gII, and they do not
affect the matter density fluctuation $\delta_m^\up{\gII}$
as shown in Sec.~\ref{sec:issue}
(see \cite{HWNO06b}).
As in Eq.~(\ref{eq:gm}), the matter density fluctuation and the expansion
perturbation are related as
\bear
\label{eq:tr}
\delta^\up{\gI}_m(\xvec,t)&=&\delta^\up{\gII}_m(\xvec,t)
-\nabla\delta^\up{\gII}_m\cdot
\int^t {dt\over a}\left({1\over a}\nabla\chi+\pv^\alpha\right)~,~~~~~\\
\kappa^\up{\gI}(\xvec,t)&=&\kappa^\up{\gII}(\xvec,t)
-\nabla\kappa^\up{\gII}\cdot\int^t {dt\over a}\left({1\over a}\nabla\chi
+\pv^\alpha\right)~.\nonumber
\enar

It is now apparent that the spatial transformation $\LL^\alpha$
in Eq.~(\ref{eq:spt}) is nothing but the spatial drift $\delta x^\alpha$
of non-relativistic matter in Eq.~(\ref{eq:sdI}). Therefore, the difference 
between the gauge choices is that while they both represent the same
proper-time hypersurface, their spatial coordinates differ in a way that
the re-labeling of the spatial coordinate in gauge choice~\gII\ violates
the mass conservation at the second order in perturbation.

Furthermore, the spatial coordinate transformation in Eq.~(\ref{eq:lpt})
can be viewed as a transformation from the Lagrangian frame (gauge
choice~\gII) to the Eulerian frame (gauge choice~\gI). As we discussed
in Sec.~\ref{sec:gauge}, the spatial drift $\delta x^\alpha$ of
non-relativistic matter in gauge choice~\gII\ vanishes, such that the
coordinate $\xvec^\up{\gII}$ in Eq.~(\ref{eq:lpt}) is identical to the
initial position~$\qvec$ 
at very early time and the spatial transformation vector
$\LL^\alpha(t,\xvec)$ corresponds to the Lagrangian displacement vector~$\DD$:
\beeq
\xvec(t,\qvec)=\qvec+\DD(t,\qvec)~,
\eneq
where $\DD$ should be distinguished from the vector perturbation $\pv_\alpha$.
In particular, the nonlinear evolution equation~(\ref{eq:pert}) at the
linear order yields
\beeq
\dot\delta=\kappa=-{\Delta\over a^2}\chi~,
\eneq
and the spatial transformation vector in Eq.~(\ref{eq:spt})
is then related to the linear-order
matter density fluctuation as
\beeq
\LL^\alpha=-\Delta^{-1}\nabla\delta_m(t,\xvec)+c(\xvec)=\DD(t,\xvec)
+\bdv{c}(\xvec)~,
\eneq
where $\bdv{c}(\xvec)$ is a scale-dependent integration constant and we ignored
the vector contribution to the transformation.

Compared to $\delta_m^\up{\gI}$ in Eq.~(\ref{eq:tr}), the matter density 
fluctuation in gauge choice~\gII\ is further compensated by the displacement
vector:
\bear
&&\left[\nabla\delta_m^\up{\gII}\cdot\DD\right](t,\kvec)\\
&&=-D^2(t)\int{d^3\qvec\over(2\pi)^3}
\left[{\qvec_1\cdot\qvec_2\over2q_1q_2}\left({q_1\over q_2}+{q_2\over q_1}
\right)\right]\delta(\qvec)\delta(\kvec-\qvec)~,\nonumber
\enar
which eliminates the dipole term in Eq.~(\ref{eq:kernI}), leading to
the kernel $F_2^\up{\gII}$ in Eq.~(\ref{eq:kernII}).

With this understanding, it is evident that gauge choices~\gI\ and~\gII\
describe the same system of irrotational non-relativistic matter flows,
but in different perspectives: Eulerian versus Lagrangian. Furthermore,
it is noted that as shown in Eq.~(\ref{eq:tr})
the matter density fluctuation in gauge choice~\gII\ is
{\it not} the one in Lagrangian perturbation theory, in which the matter
density fluctuation respects the mass and the momentum conservation.

\section{Galaxy bias in general relativity}
\label{sec:bias}
Having identified the correct temporal and spatial gauge choices for the
proper-time hypersurface, we are in a good position to discuss galaxy
bias in the context of general relativity. Galaxy bias refers to the relation
between the galaxy number density and the underlying matter distribution, 
and this relation is physically well-defined.

Due to the complexity of galaxy formation physics on small scales, 
biasing schemes are 
naturally effective descriptions, valid on large scales, which is the scale 
of our primary interest. The linear bias model (e.g., \cite{KAISE84}) is 
that the galaxy number density fluctuation $\delta_g^\up{int}$ is proportional
to the matter density fluctuation~$\delta_m$ with constant bias factor~$b$
on large scales, where the the physical galaxy number density $n_g$ is
separated into the mean and the fluctuation around it in a given coordinate
system as
\beeq
\label{eq:bng}
n_g=\bar n_g(\tp)(1+\delta_g^\up{int})~.
\eneq
In general relativity, since the matter density 
fluctuation~$\delta_m$ is gauge-dependent, the linear bias model makes little 
sense as long as the hypersurface for~$\delta_m$ remains unspecified.
In the context of general relativity, the linear bias model was recasted
\cite{BASEET11,CHLE11,JESCHI12,BRCRET12,YOHAET12} to be valid 
in the proper-time hypersurface, and in terms of our notation $\tp$~in
Eq.~(\ref{eq:bng}) is literally the proper-time and the galaxy number density
fluctuation is
\beeq
\label{eq:blin}
\delta_g^\up{int}=b~\delta_m^\up{\gI,\gII}~,
\eneq
where $\delta_m^\up{\gI}=\delta_m^\up{\gII}$ at the linear order.

The linear biasing model can be naturally extended to to the second order 
in perturbation as
\beeq
\label{eq:bsec}
\delta_g^\up{int}=b~\delta_m^\up{\gI}~,
\eneq
and using Eqs.~(\ref{eq:sec}) and~(\ref{eq:mean}) 
the galaxy number density fluctuation satisfies
\beeq
\langle\delta_g^\up{int}\rangle_\tp=0~,\qquad
\bar n_g(\tp)=\langle n_g\rangle_\tp~,
\eneq
where the ensemble average in this case is written as the spatial average
in the proper-time hypersurface. It should be
noted from Eqs.~(\ref{eq:sec}) and~(\ref{eq:mean2}) that the above relation
for $n_g$ is violated in gauge choice~\gII.

Beyond the linear order in Newtonian dynamics, local biasing models 
(e.g., \cite{SZALA88,FRGA93}) are frequently used, in which the galaxy
number density fluctuation is a nonlinear function of 
the matter density fluctuation to be expanded in a Taylor series.
At higher order, however, additional non-local terms can be included
in biasing \cite{MCRO09}, and it was shown
(e.g., \cite{MCDON06,SCJEDE13,ASBAET14,KENOET14}) 
that consistent renormalization
of galaxy bias requires the presence of
non-local derivative terms in addition to the
local terms. At the second order, the additional non-local term 
is the contraction
$s^2=s_{ij}s_{ij}$ of the gravitational tidal tensor \cite{MCRO09},
and its evidence was measured \cite{BASEET12} in simulations, where
\beeq
\label{eq:tid}
s_{ij}\equiv\nabla_i\nabla_j\phi-{1\over3}\delta^K_{ij}\delta_m=
\left[\nabla_i\nabla_j\Delta^{-1}-{1\over3}\delta^K_{ij}\right]\delta_m~,
\eneq
the normalization is $\nabla^2\phi=\delta_m$, and $\delta_{ij}^K$ is
the Kronecker delta.

To the second order, these quadratic terms in galaxy bias can be readily 
implemented to the relativistic framework, because we only need to consider 
them at the linear order:
\beeq
\label{eq:tid2}
\delta_m\RA\delta_m^\up{\gI,\gII}~,\quad 
\phi\RA{-\px\over4\pi G\bar\rho_m a^2}~,
\eneq
where the curvature potential $\px$ in the conformal Newtonian gauge is related
to the gauge choices in Table~\ref{tab:gauge} as $\px=\varphi-H\chi$ at the
linear order.\footnote{In terms of metric perturbation, the conformal 
Newtonian gauge (also known as the Poisson gauge or the longitudinal
gauge) is defined with the temporal
gauge condition $\chi=0$ and the spatial gauge condition $\gamma=0$. The
notation is written in a way that the gauge invariant variable 
$\px=\varphi-H\chi$ becomes the curvature 
perturbation~$\varphi$ in the conformal Newtonian gauge ($\chi=0$).
The scalar shear component of the normal observer is 
$\sigma_{\alpha\beta}=\chi_{,\alpha|\beta}
-\gbar_{\alpha\beta}\Delta\chi/3$, and hence the temporal gauge condition
$\chi=0$ is often called the zero-shear gauge.}

Therefore, the intrinsic fluctuation $\delta_g^\up{int}$ of the galaxy
number density in general relativity can be written 
to the second order in perturbation as
\beeq
\label{eq:bias}
\delta_g^\up{int}
=b_1~\delta_m^\up{\gI}+{1\over2}b_2\left[(\delta_m^\up{\gI})^2
-\sigma_m^2\right]+b_{s^2}\left[s^2-\langle s^2\rangle\right]~,
\eneq
where $\langle s^2\rangle=2\sigma_m^2/3$ and
it is noted that $\delta_m^\up{\gI}=\delta_m^{\up{\gI}(1)}
+\delta_m^{\up{\gI}(2)}$. Individual variables in Eq.~(\ref{eq:bias})
are gauge-invariant, and of course they can be computed in other choices
of gauge conditions, in which calculations become more involved.

Given the full second-order treatment in this paper,
we briefly touch on the third-order
galaxy bias in general relativity.
At the third order in perturbation, the additional non-local terms are
the cubic combination of the matter density fluctuation 
$\delta_m^\up{\gI}$, the gravitational tidal tensor $s_{ij}$,
and the velocity tidal tensor $t_{ij}$, such as 
$(\delta_m^\up{\gI})^3$, $s^2\delta_m^\up{\gI}$,
$s^3$, and $s_{ij}t_{ij}$ \cite{MCRO09}, where the velocity tidal tensor is,
\beeq
t_{ij}\equiv\left[\nabla_i\nabla_j\Delta^{-1}-{1\over3}\delta_{ij}^K\right]
(\theta_N-\delta_m)~,
\eneq
non-vanishing only at the second order and the normalization in Newtonian
dynamics is $\nabla^2\theta_N=\delta_m$. Noting that the perturbation variables
differ at the second order, we can readily identify the normalized
divergence as
\beeq
\label{eq:vtid}
\theta_N\RA-{1\over\HH f}~\Theta^\up{\gI}={\kappa^\up{\gI}\over Hf}~,
\eneq
and of course $\delta_m=\delta_m^\up{\gI}$
in the relativistic framework. Another non-local term in galaxy bias
that is by itself at the third order in perturbation is the scalar deviation
\cite{MCRO09}:
\beeq
\psi=\theta_N-\delta_m-{2\over7}s^2+{4\over21}\delta_m^2~,
\eneq
which vanishes up to the second-order in perturbation. We speculate that
the scalar deviation term~$\psi$ may be identified by using the 
expansion perturbation~$\kappa^\up{\gI}$ as 
\beeq
\psi\RA{\kappa^\up{\gI}\over Hf}-\delta_m^\up{\gI}-{2\over7}s^2+{4\over21}
\left(\delta_m^\up{\gI}\right)^2~,
\eneq
at the third order.

\section{DISCUSSION}
\label{sec:discussion}
We have computed the matter density fluctuation~$\dtm$ in the proper-time
hypersurface of non-relativistic matter flows to the second order in 
perturbation. It is identical to the matter density fluctuation in the
temporal comoving gauge and the spatial C-gauge (gauge choice~\gI\ in 
Table~\ref{tab:gauge}). The commonly used matter density fluctuation in the 
temporal comoving gauge and the spatial B-gauge ($N=1$, $N^\alpha=0$;
gauge choice~\gII) violates the mass conservation, while it is gauge-invariant
and also represents that in the proper-time hypersurface. 
We have provided physical understanding of each gauge condition by deriving
the geodesic path of the comoving observer, solving the nonlinear evolution
equations, and providing connections between gauge conditions. Drawing on this
finding, we have provided 
the second-order galaxy biasing in general relativity,
incorporating the nonlinear local and nonlocal terms that should be present
for consistent renormalization of galaxy bias. The second-order galaxy biasing
in this work provides an essential ingredient of the second-order relativistic
description of galaxy clustering \cite{YOZA14}.

Non-relativistic matter responds only to gravity, following geodesic path
and building up nonlinearity over time. 
When the density fluctuation becomes enormous 
$\delta_m\ge200$, the gravitationally bound objects form, and the 
trajectories of non-relativistic matter are entangled at the same time.
However, apart from these highly nonlinear regions and caustics, 
the flows of non-relativistic matter are
non-intersecting and well-defined, in particular on large scales, but
well into quasi-linear scales, which is the main reason 
the Zel'dovich approximation \cite{ZELDO70} or its variants are highly
successful in describing nonlinearity.
Therefore, on large scales, which is the scale of primary interest of
this work, the proper-time hypersurface of non-relativistic matter flows 
is physically well-defined, and the matter density fluctuation~$\dtm$ 
in the hypersurface can be computed without any ambiguity. Following the
geodesic path of non-relativistic matter, we have derived the time drift
$\delta\tp$ 
in Eq.~(\ref{eq:dev}) from the proper-time measured by the comoving observer
of non-relativistic matter and used it to provide the formula in 
Eq.~(\ref{eq:dtm}) for the matter density fluctuation~$\dtm$ 
in the proper-time hypersurface. 

Equation~(\ref{eq:dtm}) can be evaluated with any choice of gauge condition,
and it is often the case that the gauge choices in Table~\ref{tab:gauge}
are adopted to compute~$\dtm$ at the linear order using the popular
Boltzmann codes such as {\footnotesize CMBFAST} \cite{SEZA96},
{\footnotesize CAMB} \cite{LECHLA00}, and {\footnotesize CLASS} \cite{CLASS},
in which the matter density fluctuations in those gauge conditions are 
equivalent. However, at the second order in perturbations, they are different,
posing a critical question in galaxy bias ---
which matter density fluctuation represents the 
correct matter density fluctuation~$\dtm$ 
of the proper-time hypersurface that can be used 
in galaxy bias at the second order?

Gauge choice~\gII\ is the commonly used {\it comoving-synchronous} gauge,
in which there is no perturbation in the time coordinate $N=1$ 
and the off-diagonal metric component $N_\alpha=0$. With perturbations
present only in the spatial metric, the coordinates follow the
geodesic path, and the comoving observer is fixated at the spatial 
coordinates, such that the time drift of the comoving observer vanishes
$\delta\tp=0$ and 
the time coordinate in gauge choice~\gII\ is 
synchronized with the proper time of non-relativistic matter to all orders
in perturbation. Therefore, the matter density fluctuation $\delta_m^\up{\gII}$
is the matter density fluctuation $\dtm$ in the proper-time hypersurface.
Despite the presence in this gauge choice, 
the remaining spatial gauge modes, which leave 
$\delta_m^\up{\gII}$ undetermined as in Eq.~(\ref{eq:gm}), can be
projected out in $\delta_m^\up{\gII}$ {\it by hand} as in 
Eq.~(\ref{eq:kernII}).

Gauge choice~\gIII\ is the original synchronous gauge. Despite the similarity
to gauge choice~\gII, additional temporal gauge mode remains in this gauge
choice, even at the linear order, which needs to be removed by imposing
the initial condition and thereby aligning it with gauge choice~\gII.
At the second order, gauge choice~\gI\ has non-vanishing perturbation in
the time component, and the coordinate observer is on non-inertial path.
However, despite this shortcoming, the time coordinate of the
comoving observer is still {\it synchronized} with the proper-time of 
non-relativistic matter in this gauge choice ($\delta\tp=0$), and the
matter density fluctuation $\delta_m^\up{\gI}$
is the matter density fluctuation $\dtm$ in the proper-time hypersurface.
Gauge freedom is completely fixed in this gauge choice, and
$\delta_m^\up{\gI}$ in Eq.~(\ref{eq:kernI}) is different from 
$\delta_m^\up{\gII}$ at the second order.

The physical resolution to the puzzle 
comes from the gauge transformation in each gauge
condition. At the linear order, spatial gauge transformation is pure artifact
due to the spatial symmetry in the background. However, at the second order, 
spatial gauge transformation is no longer a gauge artifact, but a physical
transformation. Both gauge choices~\gI\ and~\gII\ describe the proper-time
hypersurface of non-relativistic matter, but differ in spatial coordinates
as illustrated in Eq.~(\ref{eq:lpt}). In particular, the difference in
the spatial coordinates is exactly the spatial drift of non-relativistic
matter as in Eqs.~(\ref{eq:sdI}) and~(\ref{eq:spt}) --- gauge choice~\gI\
describes the non-relativistic flows in Eulerian frame, while gauge
choice~\gII\ in Lagrangian frame.

Precisely due to this difference in spatial displacement, the matter
density fluctuations in both gauge choices differ, and one in gauge 
choice~\gII\ violates the mass conservation, arising from the spatial 
distortion in coordinates. In the rest frame of non-relativistic matter,
the proper-time is the only local observable that can be used to infer
the mean matter density of the hypersurface. However, there exists 
non-vanishing large-scale mode present
$\langle\delta_m^\up{\gII}\rangle=\sigma_m^2$ in gauge choice~\gII,
and hence $\delta_m^\up{\gII}$ cannot correctly describe the fluctuation
around the mean $\langle\rho_m\rangle_t\neq\bar\rho_m(t)$ in the proper-time
hypersurface.  Similar conclusion was drawn in \cite{HWNOET14}, in which
the one-loop matter power spectrum in gauge choice~\gII\ is computed.

With the proper identification of the matter density fluctuation in the
proper-time hypersurface of non-relativistic matter flows, it becomes
straightforward to generalize the nonlinear galaxy biasing schemes 
(e.g., \cite{SZALA88,FRGA93,MCDON06,MCRO09,BASEET12,SCJEDE13,ASBAET14}) 
in Newtonian
dynamics to those in the context of general relativity. In addition to the
linear bias term, which requires the computation of second-order matter
density fluctuation, additional nonlinear bias terms are quadratic at the
second order, and hence their individual quantities need to be evaluated
at the linear order. The additional local term~$\delta^2_m$ can be 
trivially implemented, and we have identified the additional nonlocal
term~$s^2$ from the gravitational tidal tensor as the Newtonian gauge
curvature perturbation in Eqs.~(\ref{eq:tid})
and~(\ref{eq:tid2}). The complete second-order galaxy biasing is given
in Eq.~(\ref{eq:bias}). Additional third-order terms in galaxy bias
are briefly discussed in Sec.~\ref{sec:bias}.

Recently, the second-order relativistic description of galaxy clustering
is computed by several groups \cite{YOZA14,BEMACL14b,DIDUET14}.
In \citet{BEMACL14b}, they argue that the matter density fluctuation in the
proper-time hypersurface is one $\delta_m^\up{\gII}$
in the comoving-time orthogonal gauge (gauge choice~\gII\ in our terminology).
As they correctly argue, the matter density fluctuation 
$\delta_m^\up{\gII}$ is gauge-invariant to the second order, if the
remaining gauge modes are projected out. However, as we showed in this paper,
$\delta_m^\up{\gII}$ does not properly represent the matter density fluctuation
with the mean at the local proper time, violating the mass conservation.
\citet{YOZA14} advocated the proper-time hypersurface 
for the second-order galaxy biasing scheme, and this current work 
completes the second-order relativistic description in \cite{YOZA14}
by providing the physical ground for galaxy bias.

Given the rapid development of current and future galaxy surveys and
the particular emphasis on testing gravity on large scales, theoretical
predictions need to be further improved by going beyond the linear theory,
and subtle relativistic effects in galaxy clustering need to be fully utilized 
to take advantage of precision measurements of galaxy clustering.
Equipped with the second-order galaxy biasing in this work,
the second-order general relativistic description of galaxy clustering
\cite{YOZA14} provides such natural theoretical framework, in which
further applications can build on such as the computation
of the galaxy three-point statistics for investigating the sensitivity of the
relativistic effect to the primordial non-Gaussianity and the modification
of gravity on large scales.

\acknowledgments
We acknowledge useful discussions with Zvonimir Vlah, Tobias Baldauf,
David Wands, and Toni Riotto.
J.~Y. is supported by the Swiss National Science Foundation.

\appendix   

\section{Metric perturbations and their relation to the ADM variables}
\label{app:metric}
Thorough second-order calculations are presented in \citet{NOHW04}
(see also \cite{MAWA09}). Here
we summarize the useful relations between metric perturbations and the
ADM variables that are used in the text.

Given the FRW metric perturbations in Eq.~(\ref{eq:abc}), the shift vector
and the induced spatial metric in Eq.~(\ref{eq:ADM}) are trivially matched as
\beeq
N_\alpha=-a\BB_\alpha~,\quad 
h_{\alpha\beta}=a^2\left(\gbar_{\alpha\beta}+2\CC_{\alpha\beta} \right)~.
\eneq
To the second order in perturbation, the remaining ADM variables are derived
as
\bear
N&=&1+\AA-{1\over 2}\AA^2+{1\over2}\BB^\alpha\BB_\alpha~,\\
h^{\alpha\beta}&=&{1\over a^2}\left(\gbar^{\alpha\beta}-2~\CC^{\alpha\beta}
+4~\CC^\alpha_\gamma\CC^{\beta\gamma}\right)~,\nonumber \\
N^\alpha&=&h^{\alpha\beta}N_\beta
={1\over a}\left(-\BB^\alpha+2~\BB^\beta\CC^\alpha_\beta\right)~.\nonumber
\enar
In terms of metric perturbations, the normal observer in Eq.~(\ref{eq:normal})
is 
\bear
\label{eq:nmet}
n^0&=&1-\AA+{3\over2}\AA^2-{1\over2}\BB^\alpha\BB_\alpha~,\\
n^\alpha&=&{1\over a}\left(\BB^\alpha-\AA\BB^\alpha-2~\CC^{\alpha\beta}
\BB_\beta\right)~,\nonumber
\enar
and the four velocity in Eq.~(\ref{eq:four}) is
\bear
\label{eq:umet}
\dUU&=&-\AA+{3\over 2} \AA^2 + {1\over 2}\VV^\alpha\VV_\alpha
- \VV^\alpha \BB_\alpha~,\\
\UU_0&=&-\left(1+\AA-{1\over2}\AA^2+{1\over2}\VV^\alpha\VV_\alpha\right)~,
\nonumber \\
\UU_\alpha&=&a\left(\VV_\alpha-\BB_\alpha+2\CC_{\alpha\beta}\VV^\beta\right)~,
\nonumber
\enar
where $\dUU$ is derived by the normalization condition ($\UU^a\UU_a=-1$).

The perturbation to the trace of the extrinsic curvature tensor
and the traceless part of the extrinsic curvature are 
\bear
\kappa&=&\delta K=
3H\AA-{1\over a}\left(\BB^\alpha_{\;\;|\alpha}
+\CC^{\alpha\prime}_\alpha\right)+{\AA\over a}\left({\BB^\alpha}_{|\alpha} 
+\CC^{\alpha\prime}_\alpha\right) \nonumber \\
&&
-{3\over2}H\left(3\AA^2-\BB^\alpha\BB_{\alpha}\right)
+{1\over a}\BB^\beta\left(2~\CC^\alpha_{\beta|\alpha}-\CC^\alpha_{\alpha|\beta}
\right) \nonumber \\
&&
+{2\over a}\CC^{\alpha\beta}\left(\BB_{\alpha|\beta}+\CC^\prime_{\alpha\beta}
\right)~,   \\
\sigma_{\alpha\beta}&=&a\left(\BB_{(\alpha|\beta)}+\CC^\prime_{\alpha\beta}
\right)(1-\AA)-a\BB_\gamma\left(2~\CC^\gamma_{(\alpha|\beta)}-
{\CC_{\alpha\beta}}^{|\gamma} \right)\nonumber\\
&&
- {2 \over 3} a\CC_{\alpha\beta} \left( \BB^\gamma_{\;\;|\gamma}
+\CC^{\gamma\prime}_\gamma \right)-{a\over3}\gbar_{\alpha\beta}\bigg[ 
\left({\BB^\gamma}_{|\gamma}+\CC^{\gamma\prime}_\gamma\right)
\left(1-\AA\right)\nonumber \\
&&-\BB^\gamma\left(2~\CC^\delta_{\gamma|\delta}-\CC^\delta_{\delta|\gamma} 
\right)-2~\CC^{\gamma\delta}\left(\BB_{\gamma|\delta}+
\CC^\prime_{\gamma\delta}\right) \bigg]\\
&=&-\bar K_{\alpha\beta}~,\nonumber 
\enar
and we derive the nonlinear terms in Eq.~(\ref{eq:pert}) 
\bear
\sigma_{ab}\sigma^{ab}&=&{1\over a^4}\left[\chi_{,\alpha|\beta}
\chi^{,\alpha|\beta}-{1\over3}\left(\Delta\chi\right)^2\right]
+{1\over a^2}\pv_{\alpha|\beta}\pv^{\alpha|\beta}\nonumber\\
&&
+\dot \TCC_{\alpha\beta}\dot \TCC^{\alpha\beta}
+{2\over a^2}\chi_{,\alpha|\beta}\left({1\over a}\pv^{\alpha|\beta}
+\dot\TCC^{\alpha\beta}\right) \nonumber \\
&&
+{2\over a}\pv_{\alpha|\beta}\dot\TCC^{\alpha\beta}~,\\
{1\over3}\kappa^2&+&\sigma_{ab}\sigma^{ab}=
\left({1\over a^2}\chi_{,\alpha|\beta}+{1\over a}\pv_{\alpha|\beta}
+\dot\TCC_{\alpha\beta}\right) \nonumber \\
&&
\times\left({1\over a^2}\chi^{,\alpha|\beta}+{1\over a}\pv^{\alpha|\beta}
+\dot\TCC^{\alpha\beta}\right)~.
\enar

\section{Gauge choice~\gIII\ --- The synchronous gauge}
\label{app:sync}
Despite the similarity in the metric representation to gauge choice~\gII, 
gauge choice~\gIII\ in Table~\ref{tab:gauge} leaves gauge freedoms
constrained only as in Eq.~(\ref{eq:gtIII}), and the metric perturbations
other than $\AA=\BB_\alpha=0$ are not uniquely determined by gauge 
choice~\gIII, even to the linear order in perturbation. Furthermore, 
since the comoving observer in this gauge choice differs from the normal
observer, we cannot use the nonlinear equations~(\ref{eq:pert}) derived
based on the covariant decomposition of the normal observer. It is noted
that the energy momentum tensor in Eq.~(\ref{eq:tam}) is expressed in terms
of the comoving observer and the fluid quantities would be different if they
are measured by the normal observer.

The Einstein equations in gauge choice~\gIII\ are
\bear
&&\kappa={k^2\over a^2}\chi+12\pi G\bar\rho_m a\SVV~,\\
&&\dot\kappa+2H\kappa=4\pi G\rho_m\delta_m~,\nonumber
\enar
and the conservation equations yield
\bear
&&\dot\delta_m=\kappa-{k^2\over a}\SVV~,\\
&&\dot\SVV+H\SVV=0~.\nonumber
\enar
It becomes immediately clear that the
differential equations in gauge choice~\gIII\ will become equivalent to
those in Eq.~(\ref{eq:pert}) in gauge choice~\gII, if the spatial scalar
velocity vanishes $\SVV=0$, which decays in time according to the conservation
equation. As discussed in Sec.~\ref{ssec:III}, this can be achieved by
setting $\SVV=0$ at the initial condition and thereby 
effectively assuming gauge
choice~\gII. This is the gauge choice (and the initial condition) adopted in 
the
Boltzmann codes such as {\footnotesize CMBFAST} \cite{SEZA96},
{\footnotesize CAMB} \cite{LECHLA00}, and {\footnotesize CLASS} \cite{CLASS}.

In gauge choice~\gIII, the remaining gauge modes affect the matter density
fluctuation and the expansion perturbation as
\bear
\tilde\delta_m&=&\delta_m+3\HH T=\delta_m+3Hc_1(\xvec)~,\\
\tilde\kappa&=&\kappa+\left(3\dot H+{\Delta\over a^2}\right)aT
=\kappa+3\dot Hc_1(\xvec)+{\Delta\over a^2}c_1(\xvec)~, \nonumber
\enar
where $c_1(\xvec)$ is an indeterminate scale-dependent function in 
Eq.~(\ref{eq:gtIII}). However, even in the presence of these gauge modes,
the evolution equations~(\ref{eq:pert}) for gauge choice~\gII\ can be used
at the linear order, because the gauge modes happen to be proportional to
the linear-order solutions $H$ and $\dot H$ in Eq.~(\ref{eq:pert}).
However, this accidental coincidence is absent beyond the linear order,
and one has to specifically adopt gauge choice~\gII\ to proceed further.

The metric representation in \citet{MABE95} is related to our notation as
\beeq
h_{ij}=2~\CC_{ij}~,\quad h=6~\varphi+2\Delta\gamma~,\quad \eta=-\varphi~.
\eneq

\vfill

\bibliography{bias.bbl}

\end{document}